\documentclass[fleqn,usenatbib]{mnras}

\usepackage[T1]{fontenc}
\DeclareRobustCommand{\VAN}[3]{#2}
\let\VANthebibliography\thebibliography
\def\thebibliography{\DeclareRobustCommand{\VAN}[3]{##3}\VANthebibliography}
\usepackage{soul,color}
\usepackage{graphicx}	% Including figure file
\usepackage{amsmath}	% Advanced maths commands
\usepackage{amssymb}	% Extra maths symbols
\usepackage{newtxtext,newtxmath}
\usepackage{physics}
\usepackage{booktabs}
\usepackage{multirow}

\usepackage{soul}
\newcommand{\Msun}{\, \mathrm{M}_{\odot}}

\title[Accreted stellar haloes in CDM, WDM and SIDM]{The accreted stellar haloes of Milky Way-mass galaxies as a probe of the nature of the dark matter}

\author[Victor J. Forouhar Moreno et al.]{Victor J. Forouhar Moreno$^{1,2}$\thanks{E-mail: forouhar@strw.leidenuniv.nl},
Azadeh Fattahi$^{2,3}$,
Alis J. Deason$^{2}$,
Fergus Henstridge$^{2,4}$ and
\newauthor
Alejandro Ben\'itez-Llambay$^{5}$.
\\
$^{1}$Leiden Observatory, Leiden University, Leiden 2333 CC, Netherlands\\
$^{2}$Institute for Computational Cosmology, Department of Physics, Durham University, Durham DH1 3LE, UK\\
$^{3}$The Oskar Klein Centre, Department of Physics, Stockholm University, Albanova University Center, SE-106 91 Stockholm, Sweden \\
$^{4}$Department of Physics, Lancaster University, Lancaster LA1 4YB, UK \\
$^{5}$University of Milano-Bicocca, Piazza della Scienza, 3, 20126 Milano MI, Italy
}

\date{Accepted XXX. Received YYY; in original form ZZZ}

\pubyear{2024}

\begin{document}
\label{firstpage}
\pagerange{\pageref{firstpage}--\pageref{lastpage}}
\maketitle 

\begin{abstract}
Galactic stellar haloes are largely composed of the remnants of galaxies accreted during the assembly of their host galaxies, and hence their properties reflect the mass spectrum and post-accretion evolution of their satellites. As the nature of dark matter (DM) can affect both, we explore how the properties of the accreted stellar component vary across cold (CDM), warm (WDM) and self-interacting (SIDM) models. We do this by studying accreted stellar populations around eight MW-mass haloes using cosmological hydrodynamical simulations based on the EAGLE galaxy formation model, in which we find that the accreted stellar mass remains similar across models. Contrary to WDM, which only presents minor differences relative to CDM, the distribution of accreted stars in SIDM changes significantly within $0.05R_{200}$ ($10\,\mathrm{kpc}$). The central density reduces to $\langle \rho^{\mathrm{SIDM}}_{\mathrm{exsitu}} / \rho^{\mathrm{CDM}}_{\mathrm{exsitu}} \rangle = 0.3$ and has a shallower radial dependence, with logarithmic density slopes of $\langle \alpha_{\mathrm{SIDM}} \rangle = -1.4$ \textit{vs} $\langle \alpha_{\mathrm{CDM}} \rangle = -1.7$. Additionally, stars are on more tangential orbits than their CDM counterparts, with a change in the velocity anisotropy of $\langle \Delta \beta \rangle = - 0.2$. Finally, SIDM stellar haloes have the largest number and prominence of overdensities in radius \textit{vs} radial velocity space. This is due to a combination of shorter stellar halo progenitor merging timescales and shallower host potentials, with the former resulting in less time for dynamical friction and radialisation to operate. In summary, we show that the phase-space structure of Galactic stellar haloes encode key information that can be used to distinguish and rule out different DM models.

\end{abstract}
\begin{keywords}
 dark matter, galaxies: haloes
\end{keywords}

%%%%%%%%%%%%%%%%%%%%%%%%%%%%%%%%%%%%%%%%%%%%%%%%%%

%%%%%%%%%%%%%%%%% BODY OF PAPER %%%%%%%%%%%%%%%%%%

\section{Introduction}

Structure formation within the $\Lambda$CDM model proceeds in a hierarchical fashion \citep{Davis.1985}, whereby small structures form first and then grow larger through mass accretion and mergers. This process results in the formation of virialised structures, known as dark matter haloes, that follow quasi-universal density profiles \citep{Navarro.1997} over twenty orders of magnitude in halo mass \citep{Wang.2020}. Haloes can grow in mass discretely by the accretion of neighbouring dark matter haloes, which are not necessarily disrupted. Their subsequent evolution after being accreted depends on their mass relative to their host, their orbital and structural parameters, and the nature of the dark matter itself (e.g. halos could evaporate as they merge if the dark matter has a non-zero interaction cross-section; \citealt{Kummer.2018}). Those sufficiently massive to experience dynamical friction \citep{Chandrasekhar.1943} lose orbital angular momentum and energy, eventually disrupting in the core of the host. The remaining objects are primarily affected by tidal stripping and shock heating. The loss of mass that occurs during merging and stripping can lead to the formation of streams and shells of material, leaving an imprint in phase space that can remain long after they first formed.

In the context of Milky Way (MW)-mass haloes, the hierarchical growth of structure means that stars stripped from accreted galaxies form stellar streams, shells and a dynamically hot halo of stars \citep[e.g.][]{Helmi.1999, Diemand.2005, Bullock.2005, Abadi.2006}. Stellar haloes have been detected in external galaxies \citep[e.g.][]{Radburn-Smith.2011, Gilbert.2014, Merritt.2016} and around the Milky Way. The stellar halo of our Galaxy is estimated to contain one per cent of the stellar mass \citep{Deason.2019}, and it is composed both by stars accreted from merged and stripped satellites (\textit{ex-situ}; e.g. \citealt{Bell.2008}), as well as stars that formed within the Milky Way and whose orbits were subsequently heated (\textit{in-situ}; e.g. \citealt{Purcell.2010, Cooper.2015}).

The advent of precision astrometric and spectroscopic surveys such as \textit{Gaia} \citep{Gaia.2016, GaiaDR1.2016,GaiaDR2.2018, GaiaDR3.2023}, APOGEE \citep{Majewski.2017} and H3 \citep{Conroy.2019} have helped unravel the assembly history of our Galaxy using its stellar halo. Evidence for an ancient major merger \citep{Deason.2013} has been strengthened thanks to these data \citep{Helmi.2018,Belokurov.2018}, with less massive events also being detected through their imprints in chemistry and action-space \citep[e.g.][]{Koppelman.2019, Naidu.2020}.

The prospect of leveraging the information contained within the stellar halo for galactic archaeology \citep{Eggen.1962, Searle.1978} has motivated studies seeking to establish a connection between its properties, e.g. mass, metallicity and density profiles, to how it assembled. Additionally, the large spatial extent of stellar haloes also makes them powerful tracers of the underlying gravitational potential \citep[e.g.][]{Johnston.1999}, thus helping constrain the mass \citep[e.g.][]{Deason.2021,Genina.2023}, shape \citep[e.g.][]{Bovy.2016} and distortions \citep[e.g.][]{Erkal.2021} of the dark matter halo surrounding the Milky Way.

An underlying assumption in most studies concerning the stellar halo of the Milky Way is that the dark matter is cold and collisionless. Cold dark matter (CDM), initially motivated by promising extensions to the Standard Model \citep{Ellis.1984}, became the \textit{de-facto} dark matter model as a result of the agreement between the predicted and observed properties of large scale cosmic structure \citep[e.g.][]{Cole.2005,Springel.2006, Rodriguez.2016}. However, none of the proposed particle candidates have been detected yet \citep[e.g.][]{Aprile.2018, Canepa.2019}, making viable alternatives to CDM an attractive prospect worth exploring. 

Models that differ from CDM do so primarily in their small scale predictions. 
For example, the cut-off in the matter power spectrum present in warm dark matter models (WDM) suppresses the formation of structure below the corresponding mass scale \citep{Bode.2001}. The changing power spectrum also has structural implications, such as lower halo concentrations resulting from a delay in the formation time of haloes \citep[e.g.][]{Bose.2017}. For currently viable models of WDM, these changes are relegated to the dwarf galaxy regime.

The effects of a self-interacting model of dark matter (SIDM; \citealt{Spergel.2000}) primarily affect the matter distribution where DM densities are large enough to sustain a high rate of scattering. These regions correspond to the centres of dark matter haloes, which develop flatter density profiles with a rounder configuration \citep[e.g.][]{Dave.2001} than their CDM counterparts,  so long as the cross-section is not large enough to trigger gravothermal collapse. More elaborate models also result, as is the case in WDM, in a power spectrum cut-off on small scales \citep{Vogelsberger.2015,Cyr-Racine.2016}. 

The differences arising in competing  models of DM propagate to the halos and galaxies that have been accreted by more massive systems. The suppression in the formation of low-mass haloes affects how many satellites a given Milky Way-mass host can accrete throughout its lifetime. Differences in the inner density profile are expected to affect how efficient mass-stripping due to gravitational tides is \citep[e.g.][]{Penarubia.2010}. As such, the abundance and distribution of the present day Milky Way satellites may provide a way to constrain the properties of the dark matter \citep[e.g.][]{Kennedy.2014,Lovell.2014,Newton.2021,Nadler.2021b}, modulo the uncertainties introduced by the effect of baryons and its coupling to the surrounding DM \citep{Forouhar.2022}, like supernovae-feedback on dwarf scales \citep[e.g.][]{Navarro.1996a, Pontzen.2012, DiCintio.2014,Read.2018}. 

Since the formation of the stellar halo is intimately related to the stripping and disruption of satellite galaxies, changing how many are accreted and how efficiently they are stripped of stars may result in noticeable differences in the properties of stellar haloes. For example, the suppression of the least massive satellites could affect the properties of the outer stellar halo, either in terms of its extent, mass or the amount of tidal-induced features, as these are the progenitors that dominate its outskirts \citep[e.g.][]{Fattahi.2020}. Additionally, changing the structural parameters of haloes hosting  dwarf galaxies can alter the spatial distribution and kinematics of the resulting remnant \citep{Amorisco.2017, Vasiliev.2022}.

Given the potential importance of the assumed nature of DM on the properties of stellar halos, it is surprising that few studies have explored the connection between the two in detail. Beyond stream `gapology' \citep[e.g.][]{Ibata.2002} and stellar wakes \citep[e.g.][]{Buschmann.2018}, only a limited number of recent studies have explored what effect changing the DM model has on stellar haloes. The most recent example focused on the stellar haloes around dwarf galaxies using idealised N-body simulations \citep{Deason.2022}. However, this mass scale is subject to substantial theoretical \citep[e.g.][]{Brook.2014, Read.2017, Jethwa.2018, Graus.2019, Benitez-Llambay.2020} and observational uncertainties \citep[e.g.][]{Kazantzidis.2011, Sestito.2023a, Waller.2023}. On Milky Way-mass scales, \citet{Dooley.2016} used particle tagging on N-body simulations to predict how the outer ($\geq 30~\mathrm{kpc}$) stellar halo changed between CDM and SIDM, finding virtually no differences. However, an important caveat in their study is that they only focused on the outer stellar halo, which is thought to be dominated by low mass progenitors of the stellar halo. Thus, the possibility of differences in the inner stellar halo remain, so long as the most massive progenitors (those whose debris dominate the inner regions) are affected by changes in the DM.

Addressing the question of whether the nature of dark matter systematically affects the stellar halo of Milky Way-mass galaxies is particularly important due to ongoing and upcoming surveys, such as WEAVE \citep{Jin.2023}, DESI MWS \citep{Cooper.2023}, Vera Rubin LSST \citep{Ivezic.2019} and ARRAKIHS \citep{Guzmán.2022}. These surveys will provide a wealth of data concerning the stellar halo surrounding external galaxies and the outskirts of our own. The larger sample size will provide a way of studying the statistical properties of stellar haloes around Milky Way-mass galaxies and, hence, be less dominated by the atypical assembly history of our own \citep{Evans.2020}. This is a crucial step towards using them as dark matter probes since the stochasticity in their assembly histories leads to changes comparable to those caused by alternative dark matter models \citep{Power.2016}. 

To explore how the stellar halo properties depend on the nature of the dark matter, we study eight different Milky Way-mass haloes formed within high-resolution hydrodynamical simulations based on the EAGLE model of galaxy formation. The only changes across the three cosmological simulations we consider in this work concern the assumed nature of dark matter. This allows us to study the same set of stellar haloes in cold, warm and self-interacting dark matter models, decoupling the effect of different assembly histories by matching haloes to their fiducial CDM counterparts.

We begin by introducing the simulations and galaxy formation model we used in this study, followed by the operational definitions we use to identify stellar haloes. In Section \ref{section_exsitu_mass}, we explore the present-day masses of these stellar haloes, how they compare with our own Galaxy, and whether they are sensitive to the nature of dark matter. We then proceed to investigate their spatial (\S\ref{density_stellar_halo}) and dynamical (\S\ref{kinematics_stellar_halo}) properties. Lastly, in Section \ref{progenitors_stellar_halo}, we discuss how differences in the progenitors of the stellar halos have resulted in the observed changes in their present-day properties.

\section{Simulations}

In this section, we briefly describe the simulation code, galaxy formation physics and alternative dark matter models used in our simulations. Further details are available in \citet{Forouhar.2022}.

\subsection{The galaxy formation model}

We have run our smoothed particle hydrodynamics simulations using the version of P-Gadget3 \citep{Volker.2005} that includes the galaxy formation physics of the EAGLE project \citep{Schaye.2015,Crain.2015}. The EAGLE model, which reproduces a number of observed population statistics \citep[e.g.][]{Schaller.2015,Ludlow.2017}, incorporates subgrid prescriptions for the physics relevant to galaxy formation and evolution: radiative cooling and photoheating \citep{Wiersma.2009}, star formation and evolution \citep{Schaye.2004,Schaye.2008}, stellar feedback \citep{Dalla_Vecchia.2012}, black hole seeding \citep{Springel.2005,Booth.2009} \& its subsequent growth and stochastic, thermal AGN feedback. 

The values of the parameters used in modelling these processes were set by requiring a good match to the observed $z = 0.1$ galaxy stellar mass function, the distribution of galaxy sizes and the amplitude of the central black hole mass {\em vs} stellar mass relation. For this work, we use the calibration made for the higher mass resolution version of EAGLE (RECAL in the nomenclature of \citealt{Schaye.2015}).

We simulate the evolution of structure in a periodic box of 12Mpc side-length from $z = 127$ to $z = 0$, assuming the cosmological parameter values of \citet{Planck_Collaboration.2014}. We populate it with $2 \times 512^{3}$ particles, half of which are dark matter and the rest gas particles. This corresponds to a particle mass resolution of $m_{\mathrm{DM}} = 4\times 10^{5}\,\Msun$ and $m_{\mathrm{gas}} = 8\times 10^{4} \, \Msun$, respectively. The initial DM particle distribution was generated using MUSIC \citep{Hahn.2011}.

\subsection{Alternative dark matter models}

We run two simulations in which the dark matter model differs from CDM, but whose subgrid physics and phases in the initial conditions remain the same. Any changes in the properties of stellar haloes of our simulated galaxy sample are thus primarily caused by differences driven by the nature of dark matter.

The first variation we consider is a warm dark matter model, in which we truncate the initial power spectrum of density fluctuations using the transfer function of \citet{Bode.2001}. We assume a $2.5\, \mathrm{keV}$ thermal relic particle, which corresponds to a mass scale where differences start to appear relative to CDM (its half-mode mass) of $m_{1/2} = 1.4 \times 10^{9} \Msun$. Although this mass is lighter than current observational constraints, we choose this mass to enhance differences relative to a CDM model.

The second variation in the dark matter model is through the addition of self-interactions between DM particles, modelled using the Monte-Carlo implementation of \citet{Robertson.2017}. The cross-section we use is velocity-independent and equals $\sigma_{\mathrm{SIDM}} = 10~\mathrm{cm}^{2} \mathrm{g}^{-1}$. We use the same initial conditions as those generated for the CDM simulation.

\section{Methods}

Here we summarise the methods used in \citet{Forouhar.2022} to construct the catalogue of dark matter haloes and galaxies, as well as how they are linked across time, and how spurious subhaloes are removed from the WDM simulation. We also explain how merger trees are used to identify the accreted stellar halo and its progenitor galaxies. 

\subsection{Structure finding and merger trees}

To identify dark matter haloes, we assign particles into distinct groups according to the friends-of-friends (FoF) percolation algorithm \citep{Davis.1985}. They are first found by linking every dark matter particle within 0.2 times their mean interparticle separation. The remaining particle types (gas, star and black holes) are then attached to the group of their nearest DM particle. Using the {\tt SUBFIND} algorithm \citep{Springel.2001}, the FoF groups are subdivided into candidate subhaloes by locating peaks in the smoothed density field and subjecting the enclosed particles to an iterative unbinding algorithm. Those which are self-bound and contain 20 or more particles comprise our catalogue of structures. The most massive subgroup in a given FoF group is chosen as its central galaxy, with the remaining ones labelled as satellites.

We follow the time evolution of all galaxies using their merger trees, which are built by cross-matching a subset of the most gravitationally-bound particles between consecutive time outputs \citep{Jiang.2014}. The algorithm links galaxies that temporarily disappear from the catalogues up to five consecutive time outputs, prone to occur when near the centre of a more massive object. The main progenitor branch is then found by identifying the progenitor branch with the largest integrated mass \citep{DeLucia.2007}. This choice reduces the influence that central halo switching, prone to occur during major mergers, has on the identification of the main progenitor.

\subsection{Spurious group removal}

Particle-based simulations with a resolved power spectrum cut-off result in the 
spurious fragmentation of filaments due to the discrete representation of the 
underlying density field \citep{Wang.2007}. To remove the resulting spurious subhaloes, we use the method of \citet{Lovell.2014}. First, we remove all subhaloes whose mass is lower than the mass scale where spurious structures are expected to dominate ($M_{\rm lim} = 1.4 \times 10^{8} \, \Msun$ for our setup). We then remove all remaining subhaloes whose proto-halo Lagrangian region is highly flattened, i.e. with a sphericity $s$ lower than 0.16 ($s \equiv c/a$, where $c$ and $a$ are the smallest and largest eigenvalues of the inertia tensor).

\subsection{Sample of host galaxies}

We are interested in studying the stellar haloes of Milky Way-mass galaxies, which is why we restrict our analysis to haloes of mass
$M_{200}$\footnote{$M_{200}$ is defined as the mass contained within a sphere of mean density 200 times the critical density of the universe.} at $z = 0$ in the range $[0.5 -2.5]\times 10^{12} \,
\Msun$. This is within a factor of two from recent observational estimates of the Milky Way's halo mass \citep[e.g.][]{Callingham.2019,Cautun.2020}, and results in the same selection of  eight haloes as those studied in \citet{Forouhar.2022}.

\subsection{Defining the stellar halo and its progenitors}

There are several operational definitions used to identify stellar haloes within simulations. Some rely on cuts based on the spatial \citep[e.g.][]{Monachesi.2019} and circularity distribution of stars \citep[e.g.][]{Font.2011}, and others do so based on whether the stars have been accreted from other galaxies \citep[e.g.][]{Fattahi.2020}. Choosing a particular definition reflects the questions one wants to address, which in our case concerns whether the assumed nature of dark matter affects the properties of the accreted stellar halo. As such, we define the stellar halo as being composed solely by accreted\footnote{We use accreted and ex-situ interchangeably throughout this paper.} stars.

In practice, reliably identifying which stars have formed in-situ and which have been accreted is not trivial. For instance, identifying which star particles are in-situ requires identifying the main progenitor of a given MW-mass galaxy chosen at $z = 0$. In this work we choose it to be the merger tree branch with the largest integrated mass. However, the concept of a dominant `main progenitor' becomes less well-defined at high redshifts, and major mergers can result in the misidentification of which galaxy is the most massive. Additionally, configuration space-based structure finders can temporarily miss and artificially truncate galaxies when undergoing close pericentric passages. These factors often blur the boundary between in-situ and ex-situ material.

Our selection of stars to analyse is done at redshift $z = 0$, by identifying all stellar particles within the FoF groups hosting our sample of MW-mass haloes. To avoid including in our analysis the stellar component of self-bound satellites, we further require that particles are classified as bound to either the MW central galaxy (i.e. the most massive galaxy within its FoF group) or unbound. This results in a population that includes stars formed within the MW main progenitor and stars originating from the debris of dwarf satellites accreted in the past. 

Since we are only interested in the accreted stellar component, we need to further clean the sample. We have explored two alternative ways to do so, one based on either the radial distance of stars to the Milky Way main progenitor \citep[e.g.][]{Sanderson.2018} and the other on whether stars were bound to it \citep[e.g.][]{Fattahi.2020}, both measured when the star particle formed. Although the formation time of each simulated stellar particle is precisely known, only discrete outputs in time contain information about their spatial position or structure membership. This means that, in practice, we use the snapshot immediately before the star particle birth time and hence the position, velocity and bound membership from its parent gas particle.

By comparing the above methods, we found that the spatial criterion identified some stellar shells as in-situ. This is due to several progenitor galaxies forming stars during close pericentric passages, a process likely triggered by the compression of gas by its interaction with the host \citep[e.g.][]{Genina.2019}. We therefore use the bound membership of the gas particle to identify the accreted stellar component, which correctly identified all the associated shells and streams as ex-situ.

\section{Results}

\subsection{Stellar halo masses}\label{section_exsitu_mass}

\begin{figure}
    \centering
    \includegraphics{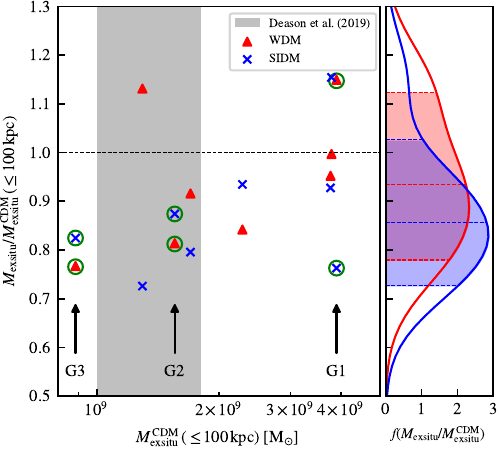}
    \caption[Ex-situ masses of Milky Way-mass haloes in different DM models]{Ex-situ stellar mass for the eight MW-mass haloes in our sample. The location along the horizontal axis indicates their value in the CDM simulation, with the mass formed in the alternative DM counterparts expressed relative to their CDM values in the vertical axis. These are indicated by the red triangles and blue crosses for WDM and SIDM, respectively. The right panel shows the distribution function of the aforementioned quantities, after applying a kernel density estimate. The shaded regions enclose the 16th and 84th percentiles of each distribution, with the median value indicated by the corresponding dashed line. The grey bands correspond to the inferred stellar halo mass of the Milky Way \citep{Deason.2019}. The three example galaxies discussed in \S\ref{kinematics_stellar_halo} (G1, G2 and G3) are highlighted using green circles and vertical arrows.}
    \label{ex-situ_mass_comparison}
\end{figure}

We begin by analysing how the accreted stellar mass varies across the galaxies in our sample, and examine if there is any dependence on the assumed model of dark matter. The ex-situ stellar masses\footnote{Measured as the total stellar mass of ex-situ star particles that are enclosed by a spherical aperture of 100~kpc placed on the centre of the galaxy.} for our sample of galaxies in the CDM simulation are shown along the horizontal axis of Fig.~\ref{ex-situ_mass_comparison}, with the median and 16th - 84th percentiles of our sample being $M^{\mathrm{CDM}}_{\mathrm{exsitu}} = 2.0^{+1.8}_{-0.7}\times 10^{9}\,\Msun$. Although roughly consistent with observational estimates for the mass of the MW stellar halo, we note that our quoted value does not include the in-situ stellar halo. Hence, the masses we measure here represent a lower limit of total mass of our simulated stellar haloes.

We further note that half of our sample has an accreted stellar halo more massive than the total mass of the Milky Way stellar halo. As our simulated Milky Way analogues are only selected based on their present-day virial mass, it is likely that the stellar halo around the galaxies in our sample formed later than counterparts whose assembly histories more closely match that of our Galaxy. There are two reasons that explain this expectation. First, galaxies hosting GE/S-like features in their stellar haloes form earlier than galaxies that do not exhibit such a radially-anisotropic component \citep{Fattahi.2019}. Hence, our mass selection will contain a higher number of `average' MW-mass haloes without a highly radially anisotropic component. Second, the stellar halo of the real Milky Way formed earlier than a mass-selected sample of simulated MW-mass galaxies \citep[e.g.][]{Deason.2019}, which also means that we might expect more massive stellar haloes (relative to the stellar mass of their host galaxy) in our sample. Indeed, the ex-situ mass fraction in our sample, defined here as the ratio of the ex-situ mass to the total stellar mass within a 30~kpc spherical aperture, is $\approx 21\%$ and hence significantly larger than the inferred $\approx 1\%$ for our Galaxy.

Shifting our focus to the impact of the assumed nature of dark matter on the properties of the accreted stellar haloes, Fig \ref{ex-situ_mass_comparison} indicates that there are up to $\approx$20\%-level differences at a fixed stellar halo mass between the different DM models. Despite a tendency toward lower accreted masses, with a median ratio and scatter of $M^{\mathrm{WDM}}_{\mathrm{exsitu}} / M^{\mathrm{CDM}}_{\mathrm{exsitu}} = 0.9^{+0.2}_{-0.2}$ and $M^{\mathrm{SIDM}}_{\mathrm{exsitu}} / M^{\mathrm{CDM}}_{\mathrm{exsitu}} = 0.8^{+0.1}_{-0.1}$, the populations are roughly consistent with those found in CDM. The relatively small differences across the various DM models explored here indicate that the nature of DM plays a minor role in setting the \textit{overall} ex-situ mass of the stellar halo. This is because the largest progenitors, which contribute most of the accreted mass, have similar stellar masses across the models, as shown in the appendix (Fig.~\ref{stellar_mass_function}). Nonetheless, it is worth remembering that we are not including the in-situ stars that have halo-like orbits in our analysis (i.e. the in-situ stellar halo). As such, our mass estimate corresponds to a lower bound of the total stellar halo mass. A future avenue to explore is whether or not larger amounts of in-situ stars have halo-like orbits in alternative dark matter models. One might expect the fraction to change owing to the shallower gravitational potentials in SIDM, making their central galaxies more susceptible to tidal perturbations.

We conclude that the total accreted mass of stellar haloes is largely insensitive to changes caused by the nature of the dark matter. The reason behind this similarity is that, as we will show explicitly in \S\ref{progenitors_stellar_halo}, neither warm nor self-interacting dark matter change the mass or abundance of the most massive stellar halo progenitors. Since these dominate the mass budget of the stellar halo, their ex-situ masses do not change substantially across models. This is perhaps not a surprising fact, as it is known that the stellar masses of the surviving population of satellite galaxies do not change significantly across DM models. Nonetheless, up until this study there had been no explicit verification of whether the stellar masses of disrupted satellites also remain relatively unchanged across different DM models. For example, one could have expected that differences in the formation time of WDM subhaloes could have reduced their stellar mass (e.g. if the time subhaloes have from when they are massive enough to form stars until they are accreted and quenched becomes significantly shorter). We have shown that, at least for the models and mass ranges we consider here, this is not the case. 

That being said, the stellar masses of the low mass end of disrupted satellites may be more sensitive to the suppression and delays in the formation of structure. If so, smaller scale structure that is part of the stellar halo may retain an imprint of the power-spectrum cut-off, e.g. in the number and properties of stellar streams. Similarly, velocity-dependent SIDM models that predict gravothermal collapse in the lowest mass subhaloes could also affect small scale stellar halo properties. Leaving aside the stellar masses, the lack of differences does not preclude changes on a spatial or dynamical level, which we explore next.

\subsection{Stellar density profiles}\label{density_stellar_halo}

As shown in earlier work \citep{Forouhar.2022}, the self-interactions between DM particles in our SIDM simulation lead to the formation of flat inner density profiles in the haloes that host dwarf and MW-mass galaxies ($5 \times 10^{9} \leq M_{200} \leq 2.5 \times 10^{12}\, \Msun$). The change on dwarf-scales accelerates the deposition of stars in the stellar halo through more efficient stripping, as the presence of a flat density core makes DM haloes less resilient to gravitational tides, relative to stepper inner density profiles \citep[e.g.][]{Penarubia.2010, Errani.2022}. On the other hand, the change on MW-scales modifies the underlying potential in which stars orbit. Thus, one might expect the spatial distribution of accreted stars to reflect these changes. 

\begin{figure}
    \centering
    \includegraphics{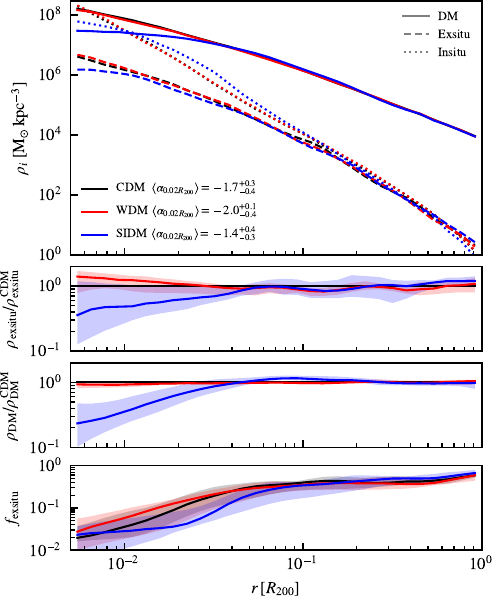}
    \caption{\textit{Top panel}: median of the 3D density profiles of the ex-situ (dashed) and in-situ (dotted) stellar components around the eight MW-mass haloes in our sample, as well as its DM component (solid). Note that the in-situ stellar density also includes the stellar disc of the galaxy. This is shown across their CDM (black), WDM (red) and SIDM (blue) variants. \textit{Middle panels}: median ratio of the ex-situ stellar and DM density at a given spherical shell across DM variants, relative to their values in the CDM counterparts. \textit{Bottom panel}: median differential ex-situ density relative to the total stellar density, as a function of distance to the centre. The distributions shown in the bottom three panels have been smoothed using a linear Savitzky–Golay filter over three consecutive bins, and the shaded regions indicate the 16th to 84th percentiles.}
    \label{ex-situ_density_profiles}
\end{figure}

We explore this in the top panel of Fig.~\ref{ex-situ_density_profiles}, which shows the median density profiles obtained from our sample of eight MW-mass haloes. We measure the profiles for each individual galaxy using 30 spherical shells whose edges are logarithmically spaced between $0.005R_{200}$ and $R_{200}$. We define the origin as the centre of mass of all dark matter particles within 5~kpc from the centre of potential of the main galaxy, as identified by {\tt SUBFIND}. The spatial offset between this centre and the halo centre of potential (or the centre of mass of the stars) is comparable to the gravitational softening of the simulation.

Focusing first on the WDM model, the density profile of accreted stars displays only minor differences relative to the CDM counterparts. The profiles are similar across the radial extent we consider here, although their densities are slightly, but systematically higher within $0.02R_{200}$. This is apparent in the second panel of Fig.~\ref{ex-situ_density_profiles}, which shows how the median value of $\rho_{\mathrm{exsitu}} / \rho^{\mathrm{CDM}}_{\mathrm{exsitu}}$ and its scatter vary across DM models as a function of distance from the centre. Overall, we find that $\rho^{\mathrm{WDM}}_{\mathrm{exsitu}} / \rho^{\mathrm{CDM}}_{\mathrm{exsitu}} = 1.4^{+0.6}_{-0.6}$ at the smallest radii we consider. The large scatter across systems makes the spatial distribution of accreted stellar halos in WDM consistent with CDM, when taken as an ensemble.

This study is not the first one to consider how the accreted stellar populations around MW-mass haloes change in a warm dark matter model. Previous work by \citet{Power.2016} found differences in the distribution of stellar material between CDM and WDM, with warmer models resulting in a stronger suppression in the \textit{total} stellar density beyond $r > 0.1 R_{200}$. In their warmest model, $m_{\mathrm{th}} = 0.5 \, \mathrm{keV}$, the density decreased by an order a magnitude, whereas it was comparable to CDM for their $m_{\mathrm{th}} = 2.0\,\mathrm{keV}$ model. Although in comparison with our simulations their galaxy sample was smaller, the mass resolution lower and the galaxy formation physics different, our findings are consistent with their coldest model. Indeed, we see no differences in the stellar density between WDM and CDM beyond $0.1 R_{200}$, as shown in the top panel of Fig.~\ref{ex-situ_density_profiles}.

Unsurprisingly, the median in-situ stellar and dark matter density profiles of the WDM model are consistent with the CDM counterparts across the whole radial range we consider. We show these density profiles in the top panel of Fig.~\ref{ex-situ_density_profiles}, which we measure using the same binning scheme as that employed for the accreted component. The similarity in the density distribution of the central galaxy and its host dark matter halo is expected. The mass scale affected by our chosen cut-off in the power spectrum is orders of magnitude lower than MW-mass scales. This means that the formation of the host DM halo is not affected by a delay in its formation, and so its structural parameters are largely the same in CDM and WDM. As the galaxy formation model is the same, the galaxies that form at its centre are very similar. 

Shifting to the SIDM model, it is apparent that the outskirts of the ex-situ density profiles are similar to those in CDM. The lack of differences in the outer stellar halo between SIDM and CDM was also highlighted by the N-body simulations used in \citet{Dooley.2016}. However, their study only explored the stellar halo density profiles beyond 30~kpc, which corresponds to approximately $(0.1 - 0.15) \times R_{\mathrm{200}}$, depending on the assumed value of $R_{200}$. Indeed, we find that the ex-situ stellar density is strongly suppressed in the central $0.05R_{200}$ in the SIDM model relative to the CDM one. The radial dependence of the accreted density in SIDM is also different to the corresponding CDM and WDM counterparts, appearing significantly flatter towards the centre. To quantify how much the slope has changed, we fit a power law, $\rho \propto r^{\alpha}$, to the ex-situ profiles within $0.02R_{200}$ for each of the eight MW-mass galaxies in our sample. We do the fitting for the stellar haloes in the CDM and SIDM models, measuring a median value of $\langle \alpha_{\mathrm{CDM}}\rangle = - 1.7^{+0.3}_{-0.4}$ and $\langle \alpha_{\mathrm{SIDM}}\rangle = -1.4^{+0.4}_{-0.3}$. Despite the large scatter in the slopes, the median accreted stellar mass profiles are systematically flatter in the central regions for stellar haloes in the SIDM model relative to CDM. Beyond the slope of the profiles, the central ($r \approx 0.01R_{200}$) ex-situ density decreases to a value of $\rho^{\mathrm{SIDM}}_{\mathrm{exsitu}} / \rho^{\mathrm{CDM}}_{\mathrm{exsitu}} = 0.3^{+0.9}_{-0.2}$ relative to the CDM counterparts, with some stellar haloes having an order of magnitude lower density.

As stellar haloes are tracers of the underlying gravitational potential field, we explore how the changes in the spatial distribution of accreted stars in SIDM  compare to the DM halo of their host galaxy. The median ratio of the DM density profiles in SIDM relative to  CDM is plotted in the second-to-last panel of Fig.~\ref{ex-situ_density_profiles}. Here we see that the suppression in the DM density at $0.01R_{200}$ ($\rho^{\mathrm{SIDM}}_{\mathrm{DM}} / \rho^{\mathrm{CDM}}_{\mathrm{DM}} = 0.2^{+0.2}_{-0.1}$) is stronger and has considerably less scatter than the suppression we measure in the accreted stellar densities. The spatial scale on which the expansion is present in both the DM and accreted stellar components is similar.

We also highlight the fact that the in-situ stellar component (which also includes the stellar disc) exhibits differences relative to their counterparts in CDM and WDM (see top panel of Fig.~\ref{ex-situ_density_profiles}). In particular, their central densities are lower than the corresponding galaxies formed in CDM and WDM, with a slight enhancement in the outskirts of the galaxy ($r \approx 0.01$ to $0.1R_{200}$). As this is likely partly driven by the shallower potential well present in the host, it will be interesting to explore in the future how the perturbative effect of mergers and fly-bys differ in the formation process of in-situ stellar haloes in SIDM cosmologies. 

Despite the fact that the ex-situ and in-situ stellar densities are suppressed in SIDM relative to CDM, the magnitude and shape of the suppression differ. Consequently, the local ex-situ mass fraction, $f_{\mathrm{exsitu}} = \rho_{\mathrm{exsitu}} / \rho_{*}$, differs in SIDM from CDM . The median values measured across our sample of eight MW-mass galaxies are shown in the bottom panel of Fig.~\ref{ex-situ_density_profiles}, where we see that the radial dependence in SIDM is substantially different to CDM and WDM. Its value is always lower in the spatial range where the halo has decreased its density compared to CDM ($r < 0.05R_{200}$), and retains a constant value within $\approx 0.02R_{200}$ instead of increasing with radius like in CDM and WDM.

We confirm that the way in which accreted stellar material is distributed within $0.05R_{200}$ varies depending on the assumed dark matter model. For WDM (SIDM), the inner regions of the accreted stellar halo are more (less) dense than in CDM, with the most extreme examples having $\rho^{\mathrm{WDM}}_{\mathrm{exsitu}} / \rho^{\mathrm{CDM}}_{\mathrm{exsitu}} = 2.56$ and $\rho^{\mathrm{SIDM}}_{\mathrm{exsitu}} / \rho^{\mathrm{CDM}}_{\mathrm{exsitu}} = 0.05$ at $0.01R_{200}$. Together with the fact that the overall ex-situ mass remains similar across models, these changes suggest a rearrangement of the accreted material, and hence how spatially concentrated it is distributed. Since the spatial distribution of stars reflects their underlying orbital distribution, these differences motivate a closer look at their velocity distributions. Additionally, phase-space signatures originating from major accretion and stripping events might differ relative to CDM, reflecting the change in the rate with which bound mass is lost and the number of accretions that result from the alternative DM variants used in this work. We explore this in the following subsection.

\subsection{Kinematics}\label{kinematics_stellar_halo}

\begin{figure}
    \centering
    \includegraphics{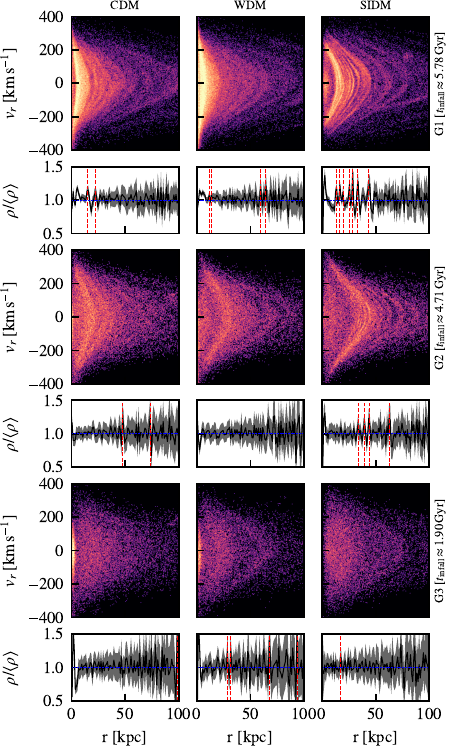}
    \caption{Radial velocity of accreted stellar particles, as a function of their radial distance to the centre of their host galaxies. The colour of each particle encodes the local ex-situ stellar density. This is shown for three $z = 0$ haloes (top to bottom) across their matched CDM, WDM and SIDM versions (left to right). The vertical ordering reflects the ease with which differences can be identified across DM models, with their IDs (G1, G2 and G3) shown on the right hand side.  The infall time of the satellite galaxy that was involved in the last major merger for each of the galaxies in this sample is indicated on the right-hand side of the corresponding row. The values are similar across DM models (see Table \ref{table_major_mergers}). The plots below each distribution show the density of ex-situ stars with $| v_{r}|\leq50\,\mathrm{km}\,\mathrm{s}^{-1}$ measured using 1~kpc radial bins, relative to the smoothed density found using a 6~kpc running mean. The solid black line and the shaded region correspond to the median and 2nd to 98th percentiles, respectively. The vertical red lines indicate peaks whose 2nd percentile values satisfy $\rho/\langle\rho\rangle > 1$, which correspond to the location of the most prominent chevrons.}
    \label{vr_vs_r_diagram_0}
\end{figure}

We now explore whether or not the velocity distribution of the ex-situ stellar halo changes across DM models. We first compare how the radial velocities of ex-situ stars, $v_{r}$, vary as a function of their distance to the centre, $r$, for three illustrative galaxies from our sample (\S\ref{representative_example_phase_space}). Following the discussion of the three illustrative galaxies, we shift our focus to the whole sample of MW-mass galaxies to investigate whether or not the velocity anisotropy of the ex-situ stars changes with the DM model in a systematic manner (\S\ref{velocity_anistropy_section}).  

\subsubsection{The phase-space and build up of the stellar halo around three illustrative galaxies}\label{representative_example_phase_space}

The three galaxies we discuss are chosen based on the fact that their $v_{r}$ vs $r$ distributions are not significantly asymmetrical, indicating that no significant disruptions of satellites have occurred in the recent past and thus provide a view of what a relaxed stellar halo looks like. From hereon, we refer to these examples as G1, G2 and G3, which is intended to represent the relative ease with which differences in the kinematics of accreted stars can be visually identified across DM models.

We show in Fig.~\ref{vr_vs_r_diagram_0} the $v_{r}$ vs $r$ distribution of the ex-situ stars for the three galaxies. The $v_{r}$ vs $r$ distributions for the other five  MW-mass haloes in our sample are shown in Fig.~\ref{vr_vs_r_rest_of_sample}, which show a greater variety in their distributions than the three examples we discuss in this section. Some galaxies, like G4, have significant asymmetries as a result of the ongoing disruption of satellites. Other galaxies, such as G5 and G6, have a mostly featureless and smooth distribution, aside from one or two overdensities in $r, v_{r}$ space (`chevrons`). The remaining two galaxies (G7 and G8) have prominent chevrons, which tend to follow the trends we identify using G1, G2 and G3 as illustrative examples.

Several chevrons are present in G1 and G2, whereas the distribution of G3 is largely smooth and featureless. Chevrons reflect apocentre pileups of stars that have been stripped from their progenitor galaxy, so their number and location encode information about the stripping history of the progenitor satellite that gave rise to them \citep[e.g.][]{Dong-Paez.2022}. The subsequent evolution is governed by interactions with the local environment \citep[e.g.][]{Davies.2023a,Davies.2023b} and Liouville's theorem, whereby chevrons progressively wind up in position and velocity space until they can no longer be distinguished from one another nor the background.

Focusing on G1 and G2, it is clear that the number of chevrons varies according to the assumed dark matter model, with WDM having the fewest and SIDM having the most. The thickness of each individual chevron also changes with the assumed dark matter model, with the thinnest chevrons being present in SIDM. In line with the spatial distribution of the accreted stellar halo, the SIDM chevrons are typically found at larger radii than in the other two models. Since this entails longer dynamical timescales, the time for the chevrons to fold and mix with the background is expected to be longer in SIDM than in WDM and CDM.

We also show in Fig.~\ref{vr_vs_r_diagram_0} how the local mass density of ex-situ stars with low radial velocities compares to the background ex-situ stellar mass density. Since stars that are part of chevrons resemble Gaussian distributions at low $v_{r}$ values \citep[e.g.][]{Donlon.2024}, the local ex-situ density at the pile-up location is higher than neighbouring radial bins. Thus, measuring the contrast between the local and background ex-situ stellar densities for low $v_{r}$ can help identify the number and location of chevrons.

In practice, we measure the local mass density of stars with $|v_{r}| \leq 50\,\mathrm{km}\,\mathrm{s}^{-1}$ using 1~kpc radial bins. We then estimate the background ex-situ density by smoothing the local densities using a running mean with a window length of 6~kpc. The smoothing is done to remove the radial dependence of the ex-situ stellar density. As the density contrast can be prone to noise, we estimate the 2nd, 50th and 98th percentile values at each radial bin after performing 20000 bootstrap resamplings of the low radial velocity stellar particles. After estimating the spread in values of the density ratio as a function of radius, we identify the location of statistically significant peaks\footnote{The number of statistically significant peaks is sensitive to the percentile threshold used to identify them, as well as the window length used to estimate the background ex-situ density. Using less stringent percentile value thresholds identifies more peaks that correspond to visually-identifiable chevrons, but also more peaks purely driven by noise fluctuations. Smoothing over smaller scales leads to more prominent peaks, including those driven by noise. Conversely, computing the average over longer spatial scales can smooth out the thinnest peaks caused by chevrons.} based on which radial bins have a 2nd percentile value above unity. If more than one consecutive radial bin satisfies this threshold, we consider them to be part of the same peak and take as its position the average of the radial bin centres. This approach results in the red vertical lines shown in the $\rho/\langle \rho\rangle$ panels of Fig.~\ref{vr_vs_r_diagram_0}. 

Beyond the inner 5~kpc of the galaxy, the location and number of statistically significant peaks in the local density contrast correlate well with those of the most prominent chevrons. For G1, the SIDM version contains the largest amount of peaks, which are typically $\approx30\%$ denser than the background in which they are embedded. Some troughs are also present at a radius immediately adjacent to the edge of a well-defined chevron. In some cases, the density contrast of troughs falls below the unity line, which is caused by the smoothing of the local density to estimate the background density. In contrast, the CDM and WDM versions of G1 do not have as many statistically significant peaks as the SIDM version, and those which are clearly present do not reach density contrasts as high as those found in SIDM. Repeating the analysis on G2 and G3, we see that the number of statistically significant density peaks reduces even further, and that the estimates get noisier due to fewer particles sampling the stellar halo.

To put these examples in perspective with respect to the rest of the sample of stellar haloes used in this study, we highlight with green circles their ex-situ masses in Fig.~\ref{ex-situ_mass_comparison}. Their masses bracket that of the overall sample, and the ordering based on the visual comparison of the differences in the number and prominence of chevrons reflects their mass ranking. Since the mass of stellar haloes typically correlate to their formation time, we proceed to estimate when the stellar haloes of G1, G2 and G3 finished accreting most of their mass.

\begin{table}
\begin{tabular}{@{}clllll@{}}
\toprule
\multicolumn{1}{l}{Event} & Model  & Mass ratio & $t_{\mathrm{infall}}$ {[}Gyr{]} & $t_{\mathrm{merge}}$ {[}Gyr{]} & $\Delta t$ {[}Gyr{]} \\ \midrule
\multirow{3}{*}{G1/M1}    & CDM    & 0.41       & 5.78                            & 9.76                           & 4.0                  \\ \cmidrule(l){3-6} 
                          & WDM    & 0.46       & 5.78                            & 10.04                          & 4.3                  \\ \cmidrule(l){3-6} 
                          & SIDM & 0.46       & 5.78                            & 8.94                           & 3.2                  \\ \cmidrule(l){2-6} 
\multirow{3}{*}{G1/M2}    & CDM    & 0.28       & 3.12                            & 5.95                           & 2.8                 \\ \cmidrule(l){3-6} 
                          & WDM    & 0.36       & 2.8                             & 6.15                           & 3.4                  \\ \cmidrule(l){3-6} 
                          & SIDM & 0.28       & 3.12                            & 5.04                           & 1.9                  \\ \midrule
\multirow{3}{*}{G2/M1}    & CDM    & 0.17       & 4.71                            & 7.93                           & 3.2                  \\ \cmidrule(l){3-6} 
                          & WDM    & 0.2        & 4.71                            & 8.17                           & 3.5                  \\ \cmidrule(l){3-6} 
                          & SIDM & 0.21       & 4.71                            & 7.68                           & 3.0                  \\ \midrule
\multirow{3}{*}{G3/M1}    & CDM    & 0.06       & 1.9                             & 3.84                           & 1.9                  \\ \cmidrule(l){2-6} 
                          & WDM    & 0.07       & 1.32                            & 3.97                           & 2.7                  \\ \cmidrule(l){2-6} 
                          & SIDM & 0.08       & 1.9                             & 2.19                           & 0.3                  \\ \bottomrule
\end{tabular}
\caption{Summary properties of the last major mergers experienced by the three examples shown in Fig. \ref{vr_vs_r_diagram_0}, identified independently across the DM variants considered in this work. The infall time, $t_{\mathrm{infall}}$, corresponds to when the merged object first became a satellite of the main progenitor. We measure the merger mass ratio in the output immediately prior to this time, before significant tidal stripping occurs. The merger is complete at $t_{\mathrm{merge}}$, which we identify as the time when the object no longer had a resolved bound component. The time elapsed between infall and disruption is shown in the rightmost column.
}
\label{table_major_mergers}
\end{table}

For this purpose, we identify for each galaxy in our sample when the its last major merger was accreted. To find the infall time of the last major merger, we follow the main branch of our G1, G2 and G3 example galaxies. We subsequently identify all galaxies that directly merged onto the main progenitor branch of either of these three example galaxies. To determine whether the accretion of a given galaxy was associated with a major merger, we compute the mass ratio between the merged galaxy and the main progenitor, $f_{\mathrm{merge}}$, at the time before the stellar halo progenitor became a satellite of the host for the first time\footnote{This definition of mass ratio avoids underestimating the merger mass ratio due to post-infall mass loss and growth of the host.}. In practice, we use the FoF group membership to identify when the stellar halo progenitor and the host main progenitor are in the same FoF group, which we use as our definition of infall. Finally, we select all luminous objects with  $f_{\mathrm{merge}} \geq 0.1$, and inspect their evolution to check whether it was indeed a major merger. If no candidates are found, we progressively lower the threshold until one is identified. This was only needed for example G3, which formed early and hence had a less trivial analysis.

The resulting infall times of the galaxy involved in the last major merger for each of the G1, G2 and G3 galaxies are $t_{\mathrm{infall}} \approx 5.78, 4.71, 1.90~\mathrm{Gyr}$, respectively. There is little to no variation in when these occur across DM models, aside from G3, which we attribute to the aforementioned difficulty associated to its early forming nature. We see that the magnitude of the differences in the abundance and prominence of chevrons across DM models correlates with the time when the last major accretion took place, with the largest changes present in galaxies that experienced later major mergers. Even if differences existed shortly after the last major merger occurred, the subsequent evolution of chevrons makes any initially existing ones mix beyond our visual identification capabilities.

This still raises the question of how the differences arise across models. To answer this, we follow the evolution of the primary progenitors of the stellar halo of G1, G2 and G3 after they are accreted. The masses and times when they are accreted are similar across models, as seen in Table ~\ref{table_major_mergers}. However, there are structural differences in the SIDM haloes resulting from self-interactions. The differences are present as flat density cores in the DM haloes hosting the progenitor galaxies of the stellar halo, as well as the velocity distribution of the stellar particles bound to them. In WDM, the main progenitors of these three stellar haloes appear to be accreted with slightly different radial velocities and timings than those in CDM and SIDM. In other words, they are at different locations in $r, v_{r}$ space than the corresponding ones in CDM and SIDM, which are more similarly located. 

\begin{figure}
    \centering
    \includegraphics{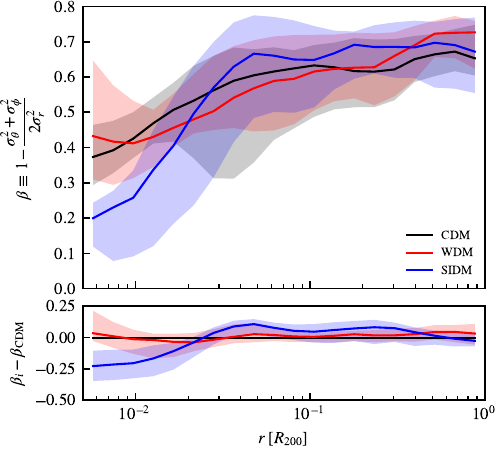}
    \caption{\textit{Top panel}: velocity anisotropy of the ex-situ stellar component, averaged across the eight MW-mass haloes in our sample in their CDM (black), WDM (red) and SIDM (blue) versions. The shaded regions indicate the 16th - 84th percentiles and the solid lines the median values. Both have been smoothed using a linear Savitzky–Golay filter over three consecutive bins. \textit{Bottom panel}: same as above, but showing the median difference in the velocity anisotropy across matched counterparts, and the associated 16th - 84th percentiles.}
    \label{velocity_anisotropy}
\end{figure}

These differences propagate into the merging time of the progenitors, which we define to be the difference in time between their accretion and disruption\footnote{Defined to occur when {\tt SUBFIND} is unable to identify them as self-bound structure during five consecutive outputs.}. As shown for the three example galaxies in Table~\ref{table_major_mergers}, they vary across models in a systematic manner: SIDM subhaloes disrupt the fastest, and the WDM ones take the longest to do so. The change in SIDM is likely due to the presence of flat density cores and subhalo evaporation, which exacerbates mass loss and accelerates their disruption. In WDM, their apocentres are larger than in CDM after the first pericentric passage. This could be due to their aforementioned (small) differences in their location in $r, v_{r}$ space relative to CDM, which are amplified after undergoing pericentre. Consequently, they remain further out and their disruption takes longer. Whether this is a systematic difference always present in accreted WDM progenitors of stellar haloes, or a difference only present in the three examples discussed here, is relegated to future work. 

The primary progenitors of the stellar haloes of G1, G2 and G3 have high mass ratios relative to the main progenitor of the MW-mass hosts (see Table~\ref{table_major_mergers}). Their post-accretion evolution is therefore driven by the interplay between dynamical friction and loss of bound mass. As their masses at infall time do not differ substantially across DM models, changing how long mergers last for could affect the integrated effect they experience from dynamical friction. Hence, the systematically different merging timescales suggest that WDM and SIDM progenitors might have experienced more or less \textit{net} dynamical friction than the corresponding CDM counterparts. This could explain why the spatial distributions end up being slightly denser in WDM and substantially less dense in SIDM, respectively. However, as noted before, the ex-situ density changes occur on the same scale as the DM ones in SIDM. This highlights the fact that the end result could be a combination of changing dynamical friction, subhalo stripping \& evaporation, and the gravitational potential of the host. Identifying which of these factors dominates is not trivial to do in the cosmological simulations we use in this work.

\subsubsection{The velocity anisotropy profile}\label{velocity_anistropy_section}

The most massive progenitors of stellar haloes generally have non-negligible masses compared to their host galaxy at infall time, as we have just illustrated using the examples of G1, G2 and G3. The most massive progenitors are therefore likely to experience a major merger, which is typically characterised by dynamical friction and orbit radialisation \citep[e.g.][]{Amorisco.2017,Vasiliev.2022}. Orbit radialisation, which results from the removal of angular momentum from the orbits of merging galaxies, makes their orbits more radial than when they were first accreted.

Motivated by the differences in how long it took for the progenitors in G1, G2 and G3 to merge across DM models, the anisotropy of accreted stars could, in principle, also change in response to varying degrees of orbit radialisation. To investigate whether this is the case, we measure the velocity anisotropy of all ex-situ stars, defined as:
\begin{equation}
    \beta = 1 - \dfrac{\sigma^{2}_{\theta} + \sigma^{2}_{\phi}}{2\sigma^{2}_{r}} \, ,
\end{equation}
across all eight MW-mass haloes in our sample.  We use a similar binning scheme as the density profiles, i.e. 20 spherical shells logarithmically spaced between $0.005R_{200}$ and $R_{200}$, and then take the median values across our whole sample. 

We measure the velocity anisotropy of the ex-situ stellar component using a galactocentric coordinate system, defined independently for each of the eight galaxies in our sample. The plane of the galaxy is defined as being perpendicular to the mass-weighted angular momentum vector of all stars within 5~kpc of its centre. We visually confirm that this procedure results in a good alignment with the stellar disc, if it is present. Two of the galaxies in our sample do not have clearly identifiable discs and have a more spheroidal stellar distribution.

We show the resulting velocity anisotropy profiles in Fig~\ref{velocity_anisotropy}, where the WDM profile is very similar to the one measured in the CDM version. In contrast, the SIDM profile exhibits clear differences with respect to CDM and WDM. We believe the differences are driven by shorter merging timescales and shallower density profiles, with the former being particularly important as it decreases the time during which orbit radialisation can occur. On the other hand, their CDM and WDM counterparts take longer to merge, giving radialisation more time to make the progenitor orbits more eccentric, and hence result in more radially supported accreted stellar haloes. 

The fact that such a difference exists is striking and highlights the potential usefulness of the Gaia-Sausage-Enceladus (GE/S) debris as a DM constraint. In other words, given current constraints on its orbital anisotropy \citep[e.g. $\beta \approx 0.8$; ][]{Belokurov.2018}, one can identify how large of an SIDM cross-section can be accommodated before radialisation becomes ineffective and no GE/S-like remnants are produced. This would provide hints on what the SIDM cross-section on scales between MW-mass and dwarf-mass scales is, corresponding to the transition region for velocity-dependent cross-sections.

Exploring the feasibility of this new potential constrain will require a more dedicated analysis than we have done here, which could take on several forms. One could be a targeted approach, whereby one identifies in a CDM simulation which MW-mass galaxies have a GE/S-like debris and see by how much its velocity anisotropy and the evolution of the GE/S progenitor differs in its SIDM counterparts. A more general approach would be to ask the question of how probable it is to find a highly radially anisotropic GE/S-like debris in different SIDM models, without having to rely on matching to CDM simulations. The benefit of not relying on CDM simulations to select the SIDM counterparts to study is that the population of galaxies and set of assembly histories that give rise to GE/S-like debris may differ between CDM and SIDM.

\begin{figure}
    \centering
    \includegraphics{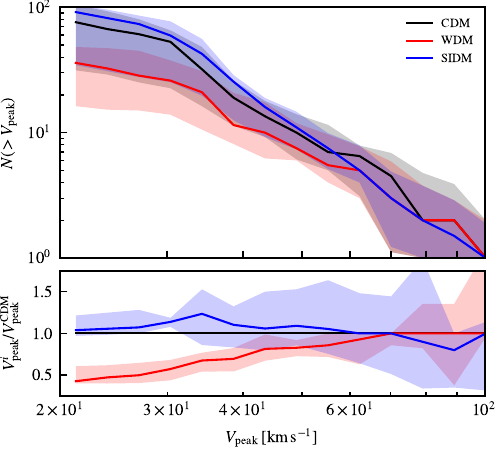}
    \caption{\textit{Top panel}: distribution of the peak maximum circular velocity of the galaxies that directly merge with the main progenitors of the eight MW-mass haloes of our sample, which we take as the building blocks of their stellar halo. This is shown across different DM models, as indicated by the top right legend. The solid lines indicate the median values across our population, with the shaded regions showing the 16th and 84th percentiles of the distribution. \textit{Bottom panel}: same as above, but for the median difference across matched counterparts. See Fig.~\ref{stellar_mass_function} for the corresponding peak stellar mass distributions.}
    \label{vpeak_distribution_orphans}
\end{figure}

In summary, we have seen that both the spatial and velocity distribution of the ex-situ stellar component depends on the assumed model of dark matter. To understand the reason why different models result in changing properties, it is crucial to investigate how the properties of its building blocks vary across models, both prior to being accreted (e.g. peak mass), and during stripping and merging (e.g. stellar deposition radii). We do so in the following subsection.

\subsection{The properties of stellar halo progenitors}\label{progenitors_stellar_halo}

\begin{figure*}
    \centering
    \includegraphics{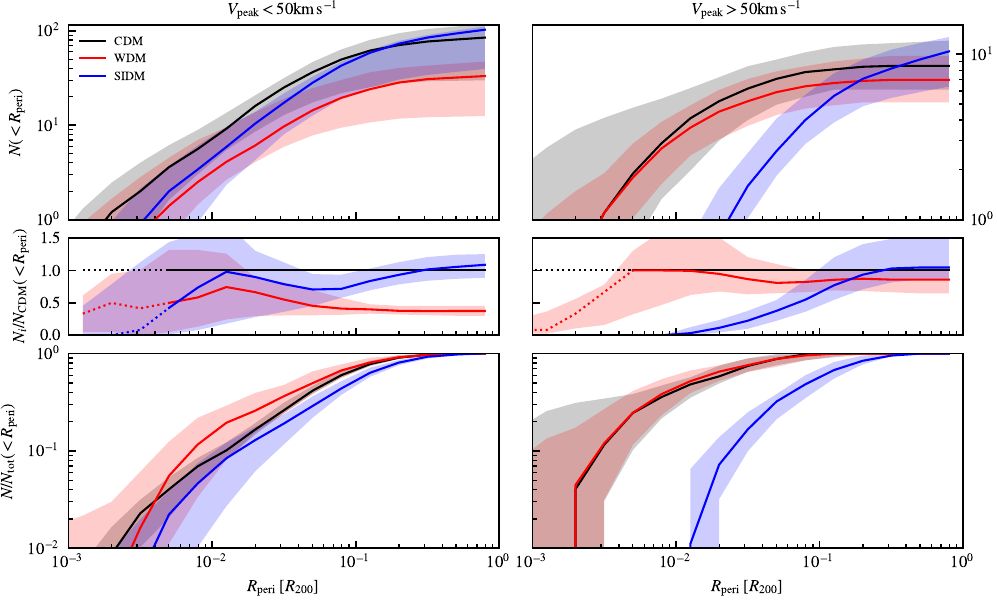}
    \caption{Cumulative distribution function of the pericentric distances of the stellar halo progenitors before they disrupted, measured for the eight MW-mass haloes in our sample. They are classified into two $V_{\mathrm{peak}}$ bins, according to whether they have values larger (right column) or lower (left column) than $50~\mathrm{km}\,\mathrm{s}^{-1}$. \textit{Top panel}: median distribution, with shaded regions bracketing the 16th and 84th percentiles. \textit{Middle panel}: same as above, but expressed relative to the value of their matched CDM counterpart. The dashed line indicates the scale at which at least one CDM counterpart reaches a value of zero, and hence where the median ratios start becoming dominated by fewer objects. \textit{Bottom panel}: same as the two panels above, but normalised to the total number of objects within $R_{200}$. All these distributions have been smoothed using a linear Savitzky–Golay filter over three consecutive bins.}
    \label{pericentric_distances_at_disruption}
\end{figure*}

The dynamical and spatial differences of present-day stellar haloes reflect the properties of accreted objects and the host assembly history. Given that we have considered the same sample of Milky Way-mass haloes, whose counterparts share the same overall assembly histories, the relative differences in the spatial and velocity distributions of stellar haloes are due to changes in the population of accreted haloes and the manner in which they contribute stars towards their build-up. We already showed this explicitly for examples G1, G2 and G3, but we now extend our analysis to see how the mass spectrum of stellar halo progenitors, as well as orbital energies, change across our whole sample of MW-mass haloes. 

To find which galaxies contributed to the build up of the stellar haloes in our sample, we identify all luminous structures whose main merger tree branch directly merges with the main progenitor branch of each MW-mass halo. We show in Fig.~\ref{vpeak_distribution_orphans} the resulting $V_{\mathrm{peak}}$\footnote{Defined as the peak value of the maximum circular velocity of a subhalo, $V_{\mathrm{circ}} = \sqrt{GM(<r)/r}$, as measured across its main progenitor branch.} function of the population we find, which is a proxy for halo mass, and hence (approximately) stellar mass.  

Focusing on the total number of disrupted objects, we note there are significant changes across dark matter models. The lowest number corresponds to the WDM model, where the suppression reflects lower numbers of galaxies forming, and hence less discrete contributions to the build-up of the stellar halo. The largest number of progenitors are present in SIDM, since it forms the same amount of structure as in CDM (their power spectra are assumed to be the same in this study), but the structural changes and scatterings with the background DM of the host result in more efficient mass loss, and hence larger numbers of disrupted subhaloes.

The differences discussed in \S\ref{density_stellar_halo} and \S\ref{kinematics_stellar_halo} concern the inner region of the stellar halo, and so are likely to be driven by its most massive progenitors. However, no significant differences are present in this mass range across models, which are all consistent within their 16th - 84th percentiles, with the most massive progenitor typically reaching $V_{\rm peak} \approx 100\,\mathrm{km} \,\mathrm{s}^{-1}$. As the models we have used here primarily affect their internal structure, or the masses and abundances of less massive haloes, the total masses of the largest contributors are unaffected, and how much mass they deposit remains similar.

However, the evolution of these massive objects prior to disruption can still vary. For example, a galaxy that experiences more tidal stripping whilst sinking towards the centre of its host, will result in stars with higher energy orbits and generally less circular orbits compared to one that is not as easily stripped. This will subsequently alter the orbital distributions of the stars, and hence their spatial and dynamical distribution. As such, we investigate this by examining the typical pericentric distance of progenitor galaxies just before they disrupted. Although galaxies deposit stars on a range of orbital energies, this metric provides an estimate on the lowest orbital energies on which stripped stars get placed onto. 

To do this, we integrate the orbits of all galaxies that are progenitors of the stellar halo in the particle potential around their corresponding MW-mass main progenitor. The potential is fit to the particle distribution using AGAMA \citep{Vasiliev.2019} and assuming a spherically symmetric potential for the stars, gas and dark matter particles. We believe that this assumption, though approximate, is sufficient to capture the major trends in the population we study here. As initial conditions, we use the relative position and velocity between the stellar halo progenitors and the MW-mass main progenitor, measured at the output when the former were last identified as self-bound by {\tt SUBFIND}. We perform this exercise for each of the eight MW-mass galaxies in our sample to obtain the corresponding distribution of pericentric distances as measured before the progenitors of their stellar haloes disrupted.

The distribution of pericentric distances just before disruption, based on the median of the distributions obtained from the eight MW-mass galaxies in our sample, is shown in Fig.~\ref{pericentric_distances_at_disruption}. Since the mass of the progenitor is an important indicator for whether they dominate the mass budget of the stellar halo, we split the distribution into two based on whether their $V_{\mathrm{peak}}$ is greater or lower than $50~\mathrm{km}\,\mathrm{s}^{-1}$. We also show these distributions normalised to the total number of stellar halo progenitors, to decouple the suppression of structure in WDM from the distribution of their disruption distances.

As the overall properties of the stellar halo are set by those with the highest $V_{\mathrm{peak}}$ values, we focus first on the $V_{\mathrm{peak}}>50~\mathrm{km}\,\mathrm{s}^{-1}$ bin. Despite a similar number of progenitors across DM models, the shape of the distributions are very different. The clearest change occurs in SIDM, where most objects disrupt at larger pericentric distances than CDM and WDM. Since the initial orbital properties are expected to be similar to CDM, the differences arise due to the subsequent orbital evolution and mass loss that the satellite galaxies experienced. As argued previously, the faster mass loss rates lead to an earlier disruption, and hence potentially lessened effects of dynamical friction and radialisation. This would generally result in their stellar remnants retaining higher orbital energy and being less radially supported than in CDM, as was found in \S\ref{density_stellar_halo} and \S\ref{kinematics_stellar_halo}. Despite this argument being particularly true for G1, G2 and G3, we have no reasons to believe this would not be the case in general.

There are essentially no changes in the WDM case, with the main one being one less galaxy typically contributing to the high $V_{\mathrm{peak}}$ distribution. Nonetheless, after normalising by the total number of progenitors we find that the distribution is the same as in CDM in the inner parts. The minor differences between the ex-situ density profiles in CDM and WDM, as well as the velocity anisotropy of the stellar halo, reflect the similarity in the distributions we find here.

Differences in the lower $V_{\mathrm{peak}}$ distribution occur primarily in the number of building blocks. As expected, the WDM counterpart is heavily suppressed relative to CDM and SIDM, and its overall distribution is more centrally concentrated. Although subhaloes disrupt at larger distances in SIDM, the difference relative to CDM is not as large as in the $V_{\mathrm{peak}} > 50~\mathrm{km}\,\mathrm{s}^{-1}$ bin. Nonetheless, this does not preclude the presence of different stellar stream properties caused by different mass loss efficiencies, which we have not explored in the present work.

\section{Conclusions}

We have used the cosmological hydrodynamical simulations presented in \citet{Forouhar.2022}, which follow the assembly of Milky Way-mass haloes across a variety of dark matter models, to explore the effect that different DM models have on the properties of accreted stellar haloes. For this purpose, we identified the accreted stellar population around eight haloes with present-day virial masses between $5\times10^{11}\,\Msun$ and $2.5\times10^{12}\,\Msun$.  Using all accreted stars as our definition of the stellar halo, our findings are as follows:

\begin{itemize}
    \item The overall accreted stellar mass remains similar across the DM models we considered in this study (Fig.~\ref{ex-situ_mass_comparison}). Since this mass is largely established by its most massive progenitors, and in particular their stellar mass, this suggests that neither DM self-interactions nor the suppression in the DM power spectrum we consider here affect their overall mass properties. This is true so long as the resulting power spectrum cut-off does not affect the formation of the main stellar halo progenitors.

    \item The spatial distribution of stellar haloes is more sensitive to the assumed model of dark matter than their ex-situ stellar mass. The differences caused by different dark matter models become the most prominent towards the inner stellar halo, with the outer stellar halo showing little-to-no differences (Fig.~\ref{ex-situ_density_profiles}). In the SIDM model, the differences appear within $0.05R_{200}$ ($\approx 10\,\mathrm{kpc}$), coinciding with the spatial scale where the DM density profile has flattened due to self-interactions. The changes manifest as a reduction of the accreted stellar density to a median value of $\langle \rho^{\mathrm{SIDM}}_{\mathrm{exsitu}} /\rho^{\mathrm{CDM}}_{\mathrm{exsitu}} \rangle = 0.3^{+0.9}_{-0.2}$ of its CDM counterparts, as well as shallower density profiles: $\langle \alpha_{\mathrm{SIDM}} \rangle= -1.4^{+0.4}_{-0.3}$ \textit{vs} $\langle \alpha_{\mathrm{CDM}} \rangle = -1.7^{+0.3}_{-0.4}$. In WDM, the profiles change in the opposite direction than SIDM, being instead slightly denser, $\langle \rho^{\mathrm{WDM}}_{\mathrm{exsitu}} /\rho^{\mathrm{CDM}}_{\mathrm{exsitu}} \rangle = 1.4^{+0.6}_{-0.6}$, and steeper, $\langle \alpha_{\mathrm{WDM}} \rangle = -2.0^{+0.1}_{-0.4}$ than the CDM counterparts. However, the changes in the WDM ex-situ density are minor compared to those observed in the SIDM model.

    \item Beyond their density profiles, the velocity distribution of accreted stars also differs across models. This results in a larger number and prominence of overdensities in $r, v_{r}$ space (Fig.~\ref{vr_vs_r_diagram_0}) and more tangentially supported orbits in SIDM than in CDM (Fig.~\ref{velocity_anisotropy}). This amounts to a typical change in the velocity anisotropy of $\langle \Delta \beta \rangle = - 0.2^{+0.1}_{-0.1}$ between SIDM and CDM in the inner $0.01R_{200}$ ($\approx 2\,\mathrm{kpc}$). The changing anisotropy could be used in conjunction with the properties of the GE/S in the Milky Way to place upper limits on the allowed SIDM cross-section. In other words, what is the maximum possible value of $\sigma_{\mathrm{SIDM}}$ that can accommodate highly radially supported ($\beta\approx0.8$) remnant of GE/S-like events. This will be explored in upcoming work.
\end{itemize}

To understand the origin behind these differences, we identified the progenitors of the stellar haloes we studied here. In this work, we defined them to be any disrupted galaxy whose descendant was the main progenitor of the haloes hosting the stellar haloes we studied.

\begin{itemize}
    \item We first discussed three representative examples that bracket the mass range of our selected sample, illustrating how the assumed dark matter model affects their merging timescales. As shown in Table~\ref{table_major_mergers}, these vary in a systematic manner, with SIDM and WDM taking the shortest and longest times to merge, respectively. 

    \item The lowest number of progenitors is present in WDM, reflecting the suppression in the amount of galaxies that form and hence accretions onto the simulated MWs. On the other hand, SIDM haloes suffer from enhanced mass loss, and so a larger number of them contribute toward the build up of stellar halo. However, these changes are relegated to the lowest mass progenitors, and hence do not contribute to the differences we discuss in this work.

    \item Given the changes in merging timescales, and hence the time that dynamical friction and radialisation have to operate, we examined how the final orbital energy of progenitors changed across models. For this purpose, we measured their pericentre values before disruption as a proxy for their final orbital energy. In doing so, we found that progenitors in SIDM have significantly larger pericentric values at disruption than CDM or WDM. Together with the expansion of the host DM halo, this constitutes another difference that can lead to the observed decrease in the central ex-situ density.
\end{itemize}

Several assumptions and limitations are present in this study. First, we have considered simulations in which the predicted differences between dark matter models are exacerbated, i.e. no flat-density core formation through gas blowouts in dwarf galaxies. As shown in \citet{Forouhar.2022}, this process can mimic the effects of SIDM on the satellite population, and could therefore similarly affect the most massive stellar halo progenitors and thus its final properties. It nonetheless remains to be seen whether baryon-driven outflows are effective at forming flat density cores for massive GE/S-like galaxies at high redshifts. Second, our comparisons between dark matter models do not include observational uncertainties. This means that the differences we identified in this work may be difficult to observe in the real Universe. As such, mock observations should be made based on these simulations, to enable a fairer comparison to real data. Third, we have used relatively extreme variations in the nature of dark matter, and so follow-up work should consider versions that are more consistent with current observational constraints.

Regarding the last point, one may indeed wonder how our findings for the SIDM model would change if we had used a velocity-dependent cross-section. Based on Fig.~\ref{vpeak_distribution_orphans}, the most massive progenitor of the stellar haloes found in our sample typically reach $V_{\rm peak} \approx 100\,\mathrm{km} \,\mathrm{s}^{-1}$. We can therefore question whether dark matter haloes with $V_{\rm max} = 100\,\mathrm{km} \,\mathrm{s}^{-1}$ have non-negligible velocity-dependent cross-sections, and hence affect their structure significantly. If so, differences in the most massive progenitors of the stellar halo will be similar to those discussed in this study, and hence result in significant differences in the kinematic and spatial distributions of ex-situ stars relative to CDM.

Several velocity-dependent cross-sections models have been used in the literature \citep[e.g.][]{Dooley.2016,Correa.2021,Correa.2024}. The models generally share similar trends at large and small relative velocities, which motivated by cluster- and dwarf-scale constraints, respectively require small and large cross-section values. However, there is up to an order of magnitude difference in the expected cross-section at $\Delta v\approx100\,\mathrm{km}\,\mathrm{s}^{-1}$, owed in part to the relative lack of constraints at these scales. Hence, whether the most significant progenitor is expected to be affected when using a velocity-dependent cross-section depends on the assumed model. Taking as an example the model used in recent cosmological hydrodynamical simulations \citep[e.g.][]{Correa.2024}, and using the maximum 1D velocity dispersion of the halo as our reference relative velocity, the expected cross-section of haloes with $V_{\mathrm{max}} = 100\,\mathrm{km}\,\mathrm{s}^{-1}$ is at least $\approx 5.25\,\mathrm{cm}^{2}\,\mathrm{g}^{-1}$.

A velocity-dependent cross-section would have other implications beyond the most massive progenitors. The high cross-section values for the lowest mass progenitor galaxies could trigger gravothermal collapse, making their central profiles denser than an NFW and hence more resilient to tides. Hence, the number and properties of tidal debris originating from them could significantly change relative to our SIDM model and CDM models. In terms of the Milky Way host dark matter halo, the low values of cross-section expected at its mass scale would likely mean that no flat density core forms. Compared to the constant cross-section we have used in this study, which does have a flat density core, the strength of the tidal field would be enhanced in the central regions.

Despite these caveats, our findings highlight a new potential avenue to constrain the nature of dark matter, based on the dynamics and distribution of the accreted stellar halo of Milky Way-mass galaxies. One approach would be to measure the spatial distribution of the stellar halo around a large sample of MW-like galaxies, and compare it to the predicted distribution across different DM models. An alternative would be to leverage the phase-space information of GE/S, by identifying simulated MW-analogues that experienced similar events in their past and measuring the final velocity anisotropy of its remnant. As changing the value of the SIDM cross-section systematically affects the merging process, and hence the anisotropy $\beta$, comparing the predicted distributions of $\beta$ to the one observed for GE/S could hint at which SIDM cross-sections are compatible with the data. 

Beyond these two approaches, one can imagine using other observables which we have not considered and could be similarly affected, such as the two-point correlation function in phase-space, clustering in action space, and metallicity gradients. Given ongoing and upcoming efforts towards studying the stellar halo Milky Way and those of external Milky Way-mass haloes, this exciting prospect warrants further investigation. 

\section*{Acknowledgements}

We thank the anonymous referee for their comments and suggestions,
which have improved the presentation and clarify of this paper. VJFM acknowledges support by NWO through the Dark Universe Science Collaboration
(OCENW.XL21.XL21.025) and the European Research Council (ERC)
through the Advanced Investigator grant DMIDAS (GA 786910) and Consolidated Grant ST/T000244/1. ABL acknowledges support by the Italian Ministry for Universities (MUR) program “Dipartimenti di Eccellenza 2023-2027” within the Centro Bicocca di Cosmologia Quantitativa (BiCoQ), and support by UNIMIB’s Fondo Di Ateneo Quota Competitiva (project 2024-ATEQC-0050). AD is supported by a Royal Society
University Research Fellowship and the Leverhulme Trust. AD and AF acknowledge support from the Science and Technology Facilities Council (STFC) [grant numbers
ST/X001075/1, ST/T000244/1]. AF has been supported by a UKRI Future Leaders Fellowship (grant no MR/T042362/1) and a Wallenberg Acadamy Fellowship. This work used the DiRAC@Durham facility managed by the Institute for Computational Cosmology on behalf of the STFC DiRAC HPC Facility (www.dirac.ac.uk). The equipment was funded by BEIS capital funding via STFC capital grants ST/K00042X/1, ST/P002293/1, ST/R002371/1 and ST/S002502/1, Durham University and STFC operations grant ST/R000832/1. DiRAC is part of the National e-Infrastructure. 
%%%%%%%%%%%%%%%%%%%%%%%%%%%%%%%%%%%%%%%%%%%%%%%%%%
\section*{Data Availability}

The data used in this study can be made available upon reasonable request to the corresponding author.

%%%%%%%%%%%%%%%%%%%% REFERENCES %%%%%%%%%%%%%%%%%%

\bibliographystyle{mnras}
\bibliography{references}

\begin{thebibliography}{}
\makeatletter
\relax
\def\mn@urlcharsother{\let\do\@makeother \do\$\do\&\do\#\do\^\do\_\do\%\do\~}
\def\mn@doi{\begingroup\mn@urlcharsother \@ifnextchar [ {\mn@doi@}
  {\mn@doi@[]}}
\def\mn@doi@[#1]#2{\def\@tempa{#1}\ifx\@tempa\@empty \href
  {http://dx.doi.org/#2} {doi:#2}\else \href {http://dx.doi.org/#2} {#1}\fi
  \endgroup}
\def\mn@eprint#1#2{\mn@eprint@#1:#2::\@nil}
\def\mn@eprint@arXiv#1{\href {http://arxiv.org/abs/#1} {{\tt arXiv:#1}}}
\def\mn@eprint@dblp#1{\href {http://dblp.uni-trier.de/rec/bibtex/#1.xml}
  {dblp:#1}}
\def\mn@eprint@#1:#2:#3:#4\@nil{\def\@tempa {#1}\def\@tempb {#2}\def\@tempc
  {#3}\ifx \@tempc \@empty \let \@tempc \@tempb \let \@tempb \@tempa \fi \ifx
  \@tempb \@empty \def\@tempb {arXiv}\fi \@ifundefined
  {mn@eprint@\@tempb}{\@tempb:\@tempc}{\expandafter \expandafter \csname
  mn@eprint@\@tempb\endcsname \expandafter{\@tempc}}}

\bibitem[\protect\citeauthoryear{{Abadi}, {Navarro}  \& {Steinmetz}}{{Abadi}
  et~al.}{2006}]{Abadi.2006}
{Abadi} M.~G.,  {Navarro} J.~F.,   {Steinmetz} M.,  2006, \mn@doi [\mnras]
  {10.1111/j.1365-2966.2005.09789.x}, \href
  {https://ui.adsabs.harvard.edu/abs/2006MNRAS.365..747A} {365, 747}

\bibitem[\protect\citeauthoryear{{Amorisco}}{{Amorisco}}{2017}]{Amorisco.2017}
{Amorisco} N.~C.,  2017, \mn@doi [\mnras] {10.1093/mnras/stw2229}, \href
  {https://ui.adsabs.harvard.edu/abs/2017MNRAS.464.2882A} {464, 2882}

\bibitem[\protect\citeauthoryear{{Aprile} et~al.,}{{Aprile}
  et~al.}{2018}]{Aprile.2018}
{Aprile} E.,  et~al., 2018, \mn@doi [\prl] {10.1103/PhysRevLett.121.111302},
  \href {https://ui.adsabs.harvard.edu/abs/2018PhRvL.121k1302A} {121, 111302}

\bibitem[\protect\citeauthoryear{{Bell} et~al.,}{{Bell}
  et~al.}{2008}]{Bell.2008}
{Bell} E.~F.,  et~al., 2008, \mn@doi [\apj] {10.1086/588032}, \href
  {https://ui.adsabs.harvard.edu/abs/2008ApJ...680..295B} {680, 295}

\bibitem[\protect\citeauthoryear{{Belokurov}, {Erkal}, {Evans}, {Koposov}  \&
  {Deason}}{{Belokurov} et~al.}{2018}]{Belokurov.2018}
{Belokurov} V.,  {Erkal} D.,  {Evans} N.~W.,  {Koposov} S.~E.,   {Deason}
  A.~J.,  2018, \mn@doi [\mnras] {10.1093/mnras/sty982}, \href
  {https://ui.adsabs.harvard.edu/abs/2018MNRAS.478..611B} {478, 611}

\bibitem[\protect\citeauthoryear{{Benitez-Llambay} \&
  {Frenk}}{{Benitez-Llambay} \& {Frenk}}{2020}]{Benitez-Llambay.2020}
{Benitez-Llambay} A.,  {Frenk} C.,  2020, \mn@doi [\mnras]
  {10.1093/mnras/staa2698}, \href
  {https://ui.adsabs.harvard.edu/abs/2020MNRAS.498.4887B} {498, 4887}

\bibitem[\protect\citeauthoryear{{Bode}, {Ostriker}  \& {Turok}}{{Bode}
  et~al.}{2001}]{Bode.2001}
{Bode} P.,  {Ostriker} J.~P.,   {Turok} N.,  2001, \mn@doi [\apj]
  {10.1086/321541}, \href
  {https://ui.adsabs.harvard.edu/abs/2001ApJ...556...93B} {556, 93}

\bibitem[\protect\citeauthoryear{{Booth} \& {Schaye}}{{Booth} \&
  {Schaye}}{2009}]{Booth.2009}
{Booth} C.~M.,  {Schaye} J.,  2009, \mn@doi [\mnras]
  {10.1111/j.1365-2966.2009.15043.x}, \href
  {https://ui.adsabs.harvard.edu/abs/2009MNRAS.398...53B} {398, 53}

\bibitem[\protect\citeauthoryear{{Bose} et~al.,}{{Bose}
  et~al.}{2017}]{Bose.2017}
{Bose} S.,  et~al., 2017, \mn@doi [\mnras] {10.1093/mnras/stw2686}, \href
  {https://ui.adsabs.harvard.edu/abs/2017MNRAS.464.4520B} {464, 4520}

\bibitem[\protect\citeauthoryear{{Bovy}, {Bahmanyar}, {Fritz}  \&
  {Kallivayalil}}{{Bovy} et~al.}{2016}]{Bovy.2016}
{Bovy} J.,  {Bahmanyar} A.,  {Fritz} T.~K.,   {Kallivayalil} N.,  2016, \mn@doi
  [\apj] {10.3847/1538-4357/833/1/31}, \href
  {https://ui.adsabs.harvard.edu/abs/2016ApJ...833...31B} {833, 31}

\bibitem[\protect\citeauthoryear{{Brook}, {Di Cintio}, {Knebe},
  {Gottl{\"o}ber}, {Hoffman}, {Yepes}  \& {Garrison-Kimmel}}{{Brook}
  et~al.}{2014}]{Brook.2014}
{Brook} C.~B.,  {Di Cintio} A.,  {Knebe} A.,  {Gottl{\"o}ber} S.,  {Hoffman}
  Y.,  {Yepes} G.,   {Garrison-Kimmel} S.,  2014, \mn@doi [\apjl]
  {10.1088/2041-8205/784/1/L14}, \href
  {https://ui.adsabs.harvard.edu/abs/2014ApJ...784L..14B} {784, L14}

\bibitem[\protect\citeauthoryear{{Bullock} \& {Johnston}}{{Bullock} \&
  {Johnston}}{2005}]{Bullock.2005}
{Bullock} J.~S.,  {Johnston} K.~V.,  2005, \mn@doi [\apj] {10.1086/497422},
  \href {https://ui.adsabs.harvard.edu/abs/2005ApJ...635..931B} {635, 931}

\bibitem[\protect\citeauthoryear{{Buschmann}, {Kopp}, {Safdi}  \&
  {Wu}}{{Buschmann} et~al.}{2018}]{Buschmann.2018}
{Buschmann} M.,  {Kopp} J.,  {Safdi} B.~R.,   {Wu} C.-L.,  2018, \mn@doi [\prl]
  {10.1103/PhysRevLett.120.211101}, \href
  {https://ui.adsabs.harvard.edu/abs/2018PhRvL.120u1101B} {120, 211101}

\bibitem[\protect\citeauthoryear{{Callingham} et~al.,}{{Callingham}
  et~al.}{2019}]{Callingham.2019}
{Callingham} T.~M.,  et~al., 2019, \mn@doi [\mnras] {10.1093/mnras/stz365},
  \href {https://ui.adsabs.harvard.edu/abs/2019MNRAS.484.5453C} {484, 5453}

\bibitem[\protect\citeauthoryear{{Canepa}}{{Canepa}}{2019}]{Canepa.2019}
{Canepa} A.,  2019, \mn@doi [Reviews in Physics] {10.1016/j.revip.2019.100033},
  \href {https://ui.adsabs.harvard.edu/abs/2019RvPhy...400033C} {4, 100033}

\bibitem[\protect\citeauthoryear{{Cautun} et~al.,}{{Cautun}
  et~al.}{2020}]{Cautun.2020}
{Cautun} M.,  et~al., 2020, \mn@doi [\mnras] {10.1093/mnras/staa1017}, \href
  {https://ui.adsabs.harvard.edu/abs/2020MNRAS.494.4291C} {494, 4291}

\bibitem[\protect\citeauthoryear{{Chandrasekhar}}{{Chandrasekhar}}{1943}]{Chandrasekhar.1943}
{Chandrasekhar} S.,  1943, \mn@doi [\apj] {10.1086/144517}, \href
  {https://ui.adsabs.harvard.edu/abs/1943ApJ....97..255C} {97, 255}

\bibitem[\protect\citeauthoryear{{Cole} et~al.,}{{Cole}
  et~al.}{2005}]{Cole.2005}
{Cole} S.,  et~al., 2005, \mn@doi [\mnras] {10.1111/j.1365-2966.2005.09318.x},
  \href {https://ui.adsabs.harvard.edu/abs/2005MNRAS.362..505C} {362, 505}

\bibitem[\protect\citeauthoryear{{Conroy} et~al.,}{{Conroy}
  et~al.}{2019}]{Conroy.2019}
{Conroy} C.,  et~al., 2019, \mn@doi [\apj] {10.3847/1538-4357/ab38b8}, \href
  {https://ui.adsabs.harvard.edu/abs/2019ApJ...883..107C} {883, 107}

\bibitem[\protect\citeauthoryear{{Cooper}, {Parry}, {Lowing}, {Cole}  \&
  {Frenk}}{{Cooper} et~al.}{2015}]{Cooper.2015}
{Cooper} A.~P.,  {Parry} O.~H.,  {Lowing} B.,  {Cole} S.,   {Frenk} C.,  2015,
  \mn@doi [\mnras] {10.1093/mnras/stv2057}, \href
  {https://ui.adsabs.harvard.edu/abs/2015MNRAS.454.3185C} {454, 3185}

\bibitem[\protect\citeauthoryear{{Cooper} et~al.,}{{Cooper}
  et~al.}{2023}]{Cooper.2023}
{Cooper} A.~P.,  et~al., 2023, \mn@doi [\apj] {10.3847/1538-4357/acb3c0}, \href
  {https://ui.adsabs.harvard.edu/abs/2023ApJ...947...37C} {947, 37}

\bibitem[\protect\citeauthoryear{{Correa}}{{Correa}}{2021}]{Correa.2021}
{Correa} C.~A.,  2021, \mn@doi [\mnras] {10.1093/mnras/stab506}, \href
  {https://ui.adsabs.harvard.edu/abs/2021MNRAS.503..920C} {503, 920}

\bibitem[\protect\citeauthoryear{{Correa}, {Schaller}, {Schaye}, {Ploeckinger},
  {Borrow}  \& {Bahe}}{{Correa} et~al.}{2024}]{Correa.2024}
{Correa} C.,  {Schaller} M.,  {Schaye} J.,  {Ploeckinger} S.,  {Borrow} J.,
  {Bahe} Y.,  2024, \mn@doi [arXiv e-prints] {10.48550/arXiv.2403.09186}, \href
  {https://ui.adsabs.harvard.edu/abs/2024arXiv240309186C} {p. arXiv:2403.09186}

\bibitem[\protect\citeauthoryear{{Crain} et~al.,}{{Crain}
  et~al.}{2015}]{Crain.2015}
{Crain} R.~A.,  et~al., 2015, \mn@doi [\mnras] {10.1093/mnras/stv725}, \href
  {https://ui.adsabs.harvard.edu/abs/2015MNRAS.450.1937C} {450, 1937}

\bibitem[\protect\citeauthoryear{{Cyr-Racine}, {Sigurdson}, {Zavala},
  {Bringmann}, {Vogelsberger}  \& {Pfrommer}}{{Cyr-Racine}
  et~al.}{2016}]{Cyr-Racine.2016}
{Cyr-Racine} F.-Y.,  {Sigurdson} K.,  {Zavala} J.,  {Bringmann} T.,
  {Vogelsberger} M.,   {Pfrommer} C.,  2016, \mn@doi [\prd]
  {10.1103/PhysRevD.93.123527}, \href
  {https://ui.adsabs.harvard.edu/abs/2016PhRvD..93l3527C} {93, 123527}

\bibitem[\protect\citeauthoryear{{Dalla Vecchia} \& {Schaye}}{{Dalla Vecchia}
  \& {Schaye}}{2012}]{Dalla_Vecchia.2012}
{Dalla Vecchia} C.,  {Schaye} J.,  2012, \mn@doi [\mnras]
  {10.1111/j.1365-2966.2012.21704.x}, \href
  {https://ui.adsabs.harvard.edu/abs/2012MNRAS.426..140D} {426, 140}

\bibitem[\protect\citeauthoryear{{Dav{\'e}}, {Spergel}, {Steinhardt}  \&
  {Wandelt}}{{Dav{\'e}} et~al.}{2001}]{Dave.2001}
{Dav{\'e}} R.,  {Spergel} D.~N.,  {Steinhardt} P.~J.,   {Wandelt} B.~D.,  2001,
  \mn@doi [\apj] {10.1086/318417}, \href
  {https://ui.adsabs.harvard.edu/abs/2001ApJ...547..574D} {547, 574}

\bibitem[\protect\citeauthoryear{{Davies}, {Vasiliev}, {Belokurov}, {Evans}  \&
  {Dillamore}}{{Davies} et~al.}{2023a}]{Davies.2023a}
{Davies} E.~Y.,  {Vasiliev} E.,  {Belokurov} V.,  {Evans} N.~W.,   {Dillamore}
  A.~M.,  2023a, \mn@doi [\mnras] {10.1093/mnras/stac3581}, \href
  {https://ui.adsabs.harvard.edu/abs/2023MNRAS.519..530D} {519, 530}

\bibitem[\protect\citeauthoryear{{Davies}, {Dillamore}, {Vasiliev}  \&
  {Belokurov}}{{Davies} et~al.}{2023b}]{Davies.2023b}
{Davies} E.~Y.,  {Dillamore} A.~M.,  {Vasiliev} E.,   {Belokurov} V.,  2023b,
  \mn@doi [\mnras] {10.1093/mnrasl/slad017}, \href
  {https://ui.adsabs.harvard.edu/abs/2023MNRAS.521L..24D} {521, L24}

\bibitem[\protect\citeauthoryear{{Davis}, {Efstathiou}, {Frenk}  \&
  {White}}{{Davis} et~al.}{1985}]{Davis.1985}
{Davis} M.,  {Efstathiou} G.,  {Frenk} C.~S.,   {White} S.~D.~M.,  1985,
  \mn@doi [\apj] {10.1086/163168}, \href
  {https://ui.adsabs.harvard.edu/abs/1985ApJ...292..371D} {292, 371}

\bibitem[\protect\citeauthoryear{{De Lucia} \& {Blaizot}}{{De Lucia} \&
  {Blaizot}}{2007}]{DeLucia.2007}
{De Lucia} G.,  {Blaizot} J.,  2007, \mn@doi [\mnras]
  {10.1111/j.1365-2966.2006.11287.x}, \href
  {https://ui.adsabs.harvard.edu/abs/2007MNRAS.375....2D} {375, 2}

\bibitem[\protect\citeauthoryear{{Deason}, {Belokurov}, {Evans}  \&
  {Johnston}}{{Deason} et~al.}{2013}]{Deason.2013}
{Deason} A.~J.,  {Belokurov} V.,  {Evans} N.~W.,   {Johnston} K.~V.,  2013,
  \mn@doi [\apj] {10.1088/0004-637X/763/2/113}, \href
  {https://ui.adsabs.harvard.edu/abs/2013ApJ...763..113D} {763, 113}

\bibitem[\protect\citeauthoryear{{Deason}, {Belokurov}  \& {Sanders}}{{Deason}
  et~al.}{2019}]{Deason.2019}
{Deason} A.~J.,  {Belokurov} V.,   {Sanders} J.~L.,  2019, \mn@doi [\mnras]
  {10.1093/mnras/stz2793}, \href
  {https://ui.adsabs.harvard.edu/abs/2019MNRAS.490.3426D} {490, 3426}

\bibitem[\protect\citeauthoryear{{Deason} et~al.,}{{Deason}
  et~al.}{2021}]{Deason.2021}
{Deason} A.~J.,  et~al., 2021, \mn@doi [\mnras] {10.1093/mnras/staa3984}, \href
  {https://ui.adsabs.harvard.edu/abs/2021MNRAS.501.5964D} {501, 5964}

\bibitem[\protect\citeauthoryear{{Deason}, {Bose}, {Fattahi}, {Amorisco},
  {Hellwing}  \& {Frenk}}{{Deason} et~al.}{2022}]{Deason.2022}
{Deason} A.~J.,  {Bose} S.,  {Fattahi} A.,  {Amorisco} N.~C.,  {Hellwing} W.,
  {Frenk} C.~S.,  2022, \mn@doi [\mnras] {10.1093/mnras/stab3524}, \href
  {https://ui.adsabs.harvard.edu/abs/2022MNRAS.511.4044D} {511, 4044}

\bibitem[\protect\citeauthoryear{{Di Cintio}, {Brook}, {Dutton}, {Macci{\`o}},
  {Stinson}  \& {Knebe}}{{Di Cintio} et~al.}{2014}]{DiCintio.2014}
{Di Cintio} A.,  {Brook} C.~B.,  {Dutton} A.~A.,  {Macci{\`o}} A.~V.,
  {Stinson} G.~S.,   {Knebe} A.,  2014, \mn@doi [\mnras]
  {10.1093/mnras/stu729}, \href
  {https://ui.adsabs.harvard.edu/abs/2014MNRAS.441.2986D} {441, 2986}

\bibitem[\protect\citeauthoryear{{Diemand}, {Madau}  \& {Moore}}{{Diemand}
  et~al.}{2005}]{Diemand.2005}
{Diemand} J.,  {Madau} P.,   {Moore} B.,  2005, \mn@doi [\mnras]
  {10.1111/j.1365-2966.2005.09604.x}, \href
  {https://ui.adsabs.harvard.edu/abs/2005MNRAS.364..367D} {364, 367}

\bibitem[\protect\citeauthoryear{{Dong-P{\'a}ez}, {Vasiliev}  \&
  {Evans}}{{Dong-P{\'a}ez} et~al.}{2022}]{Dong-Paez.2022}
{Dong-P{\'a}ez} C.~A.,  {Vasiliev} E.,   {Evans} N.~W.,  2022, \mn@doi [\mnras]
  {10.1093/mnras/stab3361}, \href
  {https://ui.adsabs.harvard.edu/abs/2022MNRAS.510..230D} {510, 230}

\bibitem[\protect\citeauthoryear{{Donlon}, {Newberg}, {Sanderson}, {Bregou},
  {Horta}, {Arora}  \& {Panithanpaisal}}{{Donlon} et~al.}{2024}]{Donlon.2024}
{Donlon} T.,  {Newberg} H.~J.,  {Sanderson} R.,  {Bregou} E.,  {Horta} D.,
  {Arora} A.,   {Panithanpaisal} N.,  2024, \mn@doi [\mnras]
  {10.1093/mnras/stae1264}, \href
  {https://ui.adsabs.harvard.edu/abs/2024MNRAS.531.1422D} {531, 1422}

\bibitem[\protect\citeauthoryear{{Dooley}, {Peter}, {Vogelsberger}, {Zavala}
  \& {Frebel}}{{Dooley} et~al.}{2016}]{Dooley.2016}
{Dooley} G.~A.,  {Peter} A. H.~G.,  {Vogelsberger} M.,  {Zavala} J.,   {Frebel}
  A.,  2016, \mn@doi [\mnras] {10.1093/mnras/stw1309}, \href
  {https://ui.adsabs.harvard.edu/abs/2016MNRAS.461..710D} {461, 710}

\bibitem[\protect\citeauthoryear{{Eggen}, {Lynden-Bell}  \& {Sandage}}{{Eggen}
  et~al.}{1962}]{Eggen.1962}
{Eggen} O.~J.,  {Lynden-Bell} D.,   {Sandage} A.~R.,  1962, \mn@doi [\apj]
  {10.1086/147433}, \href
  {https://ui.adsabs.harvard.edu/abs/1962ApJ...136..748E} {136, 748}

\bibitem[\protect\citeauthoryear{{Ellis}, {Hagelin}, {Nanopoulos}, {Olive}  \&
  {Srednicki}}{{Ellis} et~al.}{1984}]{Ellis.1984}
{Ellis} J.,  {Hagelin} J.~S.,  {Nanopoulos} D.~V.,  {Olive} K.,   {Srednicki}
  M.,  1984, \mn@doi [Nuclear Physics B] {10.1016/0550-3213(84)90461-9}, \href
  {https://ui.adsabs.harvard.edu/abs/1984NuPhB.238..453E} {238, 453}

\bibitem[\protect\citeauthoryear{{Erkal} et~al.,}{{Erkal}
  et~al.}{2021}]{Erkal.2021}
{Erkal} D.,  et~al., 2021, \mn@doi [\mnras] {10.1093/mnras/stab1828}, \href
  {https://ui.adsabs.harvard.edu/abs/2021MNRAS.506.2677E} {506, 2677}

\bibitem[\protect\citeauthoryear{Errani, Navarro, Peñarrubia, Famaey  \&
  Ibata}{Errani et~al.}{2022}]{Errani.2022}
Errani R.,  Navarro J.~F.,  Peñarrubia J.,  Famaey B.,   Ibata R.,  2022,
  \mn@doi [Monthly Notices of the Royal Astronomical Society]
  {10.1093/mnras/stac3499}, 519, 384

\bibitem[\protect\citeauthoryear{{Evans}, {Fattahi}, {Deason}  \&
  {Frenk}}{{Evans} et~al.}{2020}]{Evans.2020}
{Evans} T.~A.,  {Fattahi} A.,  {Deason} A.~J.,   {Frenk} C.~S.,  2020, \mn@doi
  [\mnras] {10.1093/mnras/staa2202}, \href
  {https://ui.adsabs.harvard.edu/abs/2020MNRAS.497.4311E} {497, 4311}

\bibitem[\protect\citeauthoryear{{Fattahi} et~al.,}{{Fattahi}
  et~al.}{2019}]{Fattahi.2019}
{Fattahi} A.,  et~al., 2019, \mn@doi [\mnras] {10.1093/mnras/stz159}, \href
  {https://ui.adsabs.harvard.edu/abs/2019MNRAS.484.4471F} {484, 4471}

\bibitem[\protect\citeauthoryear{{Fattahi} et~al.,}{{Fattahi}
  et~al.}{2020}]{Fattahi.2020}
{Fattahi} A.,  et~al., 2020, \mn@doi [\mnras] {10.1093/mnras/staa2221}, \href
  {https://ui.adsabs.harvard.edu/abs/2020MNRAS.497.4459F} {497, 4459}

\bibitem[\protect\citeauthoryear{{Font}, {McCarthy}, {Crain}, {Theuns},
  {Schaye}, {Wiersma}  \& {Dalla Vecchia}}{{Font} et~al.}{2011}]{Font.2011}
{Font} A.~S.,  {McCarthy} I.~G.,  {Crain} R.~A.,  {Theuns} T.,  {Schaye} J.,
  {Wiersma} R.~P.~C.,   {Dalla Vecchia} C.,  2011, \mn@doi [\mnras]
  {10.1111/j.1365-2966.2011.19227.x}, \href
  {https://ui.adsabs.harvard.edu/abs/2011MNRAS.416.2802F} {416, 2802}

\bibitem[\protect\citeauthoryear{{Forouhar Moreno}, {Ben{\'\i}tez-Llambay},
  {Cole}  \& {Frenk}}{{Forouhar Moreno} et~al.}{2022}]{Forouhar.2022}
{Forouhar Moreno} V.~J.,  {Ben{\'\i}tez-Llambay} A.,  {Cole} S.,   {Frenk} C.,
  2022, \mn@doi [\mnras] {10.1093/mnras/stac3062}, \href
  {https://ui.adsabs.harvard.edu/abs/2022MNRAS.517.5627F} {517, 5627}

\bibitem[\protect\citeauthoryear{{Gaia Collaboration} et~al.,}{{Gaia
  Collaboration} et~al.}{2016a}]{Gaia.2016}
{Gaia Collaboration} et~al., 2016a, \mn@doi [\aap]
  {10.1051/0004-6361/201629272}, \href
  {https://ui.adsabs.harvard.edu/abs/2016A&A...595A...1G} {595, A1}

\bibitem[\protect\citeauthoryear{{Gaia Collaboration} et~al.,}{{Gaia
  Collaboration} et~al.}{2016b}]{GaiaDR1.2016}
{Gaia Collaboration} et~al., 2016b, \mn@doi [\aap]
  {10.1051/0004-6361/201629512}, \href
  {https://ui.adsabs.harvard.edu/abs/2016A&A...595A...2G} {595, A2}

\bibitem[\protect\citeauthoryear{{Gaia Collaboration} et~al.,}{{Gaia
  Collaboration} et~al.}{2018}]{GaiaDR2.2018}
{Gaia Collaboration} et~al., 2018, \mn@doi [\aap]
  {10.1051/0004-6361/201833051}, \href
  {https://ui.adsabs.harvard.edu/abs/2018A&A...616A...1G} {616, A1}

\bibitem[\protect\citeauthoryear{{Gaia Collaboration} et~al.,}{{Gaia
  Collaboration} et~al.}{2023}]{GaiaDR3.2023}
{Gaia Collaboration} et~al., 2023, \mn@doi [\aap]
  {10.1051/0004-6361/202243940}, \href
  {https://ui.adsabs.harvard.edu/abs/2023A&A...674A...1G} {674, A1}

\bibitem[\protect\citeauthoryear{{Genina}, {Frenk}, {Ben{\'\i}tez-Llambay},
  {Cole}, {Navarro}, {Oman}  \& {Fattahi}}{{Genina} et~al.}{2019}]{Genina.2019}
{Genina} A.,  {Frenk} C.~S.,  {Ben{\'\i}tez-Llambay} A.,  {Cole} S.,  {Navarro}
  J.~F.,  {Oman} K.~A.,   {Fattahi} A.,  2019, \mn@doi [\mnras]
  {10.1093/mnras/stz1852}, \href
  {https://ui.adsabs.harvard.edu/abs/2019MNRAS.488.2312G} {488, 2312}

\bibitem[\protect\citeauthoryear{{Genina}, {Deason}  \& {Frenk}}{{Genina}
  et~al.}{2023}]{Genina.2023}
{Genina} A.,  {Deason} A.~J.,   {Frenk} C.~S.,  2023, \mn@doi [\mnras]
  {10.1093/mnras/stad397}, \href
  {https://ui.adsabs.harvard.edu/abs/2023MNRAS.520.3767G} {520, 3767}

\bibitem[\protect\citeauthoryear{{Gilbert} et~al.,}{{Gilbert}
  et~al.}{2014}]{Gilbert.2014}
{Gilbert} K.~M.,  et~al., 2014, \mn@doi [\apj] {10.1088/0004-637X/796/2/76},
  \href {https://ui.adsabs.harvard.edu/abs/2014ApJ...796...76G} {796, 76}

\bibitem[\protect\citeauthoryear{{Graus}, {Bullock}, {Kelley},
  {Boylan-Kolchin}, {Garrison-Kimmel}  \& {Qi}}{{Graus}
  et~al.}{2019}]{Graus.2019}
{Graus} A.~S.,  {Bullock} J.~S.,  {Kelley} T.,  {Boylan-Kolchin} M.,
  {Garrison-Kimmel} S.,   {Qi} Y.,  2019, \mn@doi [\mnras]
  {10.1093/mnras/stz1992}, \href
  {https://ui.adsabs.harvard.edu/abs/2019MNRAS.488.4585G} {488, 4585}

\bibitem[\protect\citeauthoryear{Guzm\'an~et al.}{Guzm\'an~et
  al.}{2022}]{Guzmán.2022}
Guzm\'an~et al. R.,  2022, ARRAKIHS Proposal

\bibitem[\protect\citeauthoryear{{Hahn} \& {Abel}}{{Hahn} \&
  {Abel}}{2011}]{Hahn.2011}
{Hahn} O.,  {Abel} T.,  2011, \mn@doi [\mnras]
  {10.1111/j.1365-2966.2011.18820.x}, \href
  {https://ui.adsabs.harvard.edu/abs/2011MNRAS.415.2101H} {415, 2101}

\bibitem[\protect\citeauthoryear{{Helmi} \& {White}}{{Helmi} \&
  {White}}{1999}]{Helmi.1999}
{Helmi} A.,  {White} S. D.~M.,  1999, \mn@doi [\mnras]
  {10.1046/j.1365-8711.1999.02616.x}, \href
  {https://ui.adsabs.harvard.edu/abs/1999MNRAS.307..495H} {307, 495}

\bibitem[\protect\citeauthoryear{{Helmi}, {Babusiaux}, {Koppelman}, {Massari},
  {Veljanoski}  \& {Brown}}{{Helmi} et~al.}{2018}]{Helmi.2018}
{Helmi} A.,  {Babusiaux} C.,  {Koppelman} H.~H.,  {Massari} D.,  {Veljanoski}
  J.,   {Brown} A. G.~A.,  2018, \mn@doi [\nat] {10.1038/s41586-018-0625-x},
  \href {https://ui.adsabs.harvard.edu/abs/2018Natur.563...85H} {563, 85}

\bibitem[\protect\citeauthoryear{{Ibata}, {Lewis}, {Irwin}  \& {Quinn}}{{Ibata}
  et~al.}{2002}]{Ibata.2002}
{Ibata} R.~A.,  {Lewis} G.~F.,  {Irwin} M.~J.,   {Quinn} T.,  2002, \mn@doi
  [\mnras] {10.1046/j.1365-8711.2002.05358.x}, \href
  {https://ui.adsabs.harvard.edu/abs/2002MNRAS.332..915I} {332, 915}

\bibitem[\protect\citeauthoryear{{Ivezi{\'c}} et~al.,}{{Ivezi{\'c}}
  et~al.}{2019}]{Ivezic.2019}
{Ivezi{\'c}} {\v{Z}}.,  et~al., 2019, \mn@doi [\apj]
  {10.3847/1538-4357/ab042c}, \href
  {https://ui.adsabs.harvard.edu/abs/2019ApJ...873..111I} {873, 111}

\bibitem[\protect\citeauthoryear{{Jethwa}, {Erkal}  \& {Belokurov}}{{Jethwa}
  et~al.}{2018}]{Jethwa.2018}
{Jethwa} P.,  {Erkal} D.,   {Belokurov} V.,  2018, \mn@doi [\mnras]
  {10.1093/mnras/stx2330}, \href
  {https://ui.adsabs.harvard.edu/abs/2018MNRAS.473.2060J} {473, 2060}

\bibitem[\protect\citeauthoryear{{Jiang}, {Helly}, {Cole}  \& {Frenk}}{{Jiang}
  et~al.}{2014}]{Jiang.2014}
{Jiang} L.,  {Helly} J.~C.,  {Cole} S.,   {Frenk} C.~S.,  2014, \mn@doi
  [\mnras] {10.1093/mnras/stu390}, \href
  {https://ui.adsabs.harvard.edu/abs/2014MNRAS.440.2115J} {440, 2115}

\bibitem[\protect\citeauthoryear{{Jin} et~al.,}{{Jin} et~al.}{2023}]{Jin.2023}
{Jin} S.,  et~al., 2023, \mn@doi [\mnras] {10.1093/mnras/stad557}, \href
  {https://ui.adsabs.harvard.edu/abs/2023MNRAS.tmp..715J} {}

\bibitem[\protect\citeauthoryear{{Johnston}, {Zhao}, {Spergel}  \&
  {Hernquist}}{{Johnston} et~al.}{1999}]{Johnston.1999}
{Johnston} K.~V.,  {Zhao} H.,  {Spergel} D.~N.,   {Hernquist} L.,  1999,
  \mn@doi [\apjl] {10.1086/311876}, \href
  {https://ui.adsabs.harvard.edu/abs/1999ApJ...512L.109J} {512, L109}

\bibitem[\protect\citeauthoryear{{Kazantzidis}, {{\L}okas}, {Callegari},
  {Mayer}  \& {Moustakas}}{{Kazantzidis} et~al.}{2011}]{Kazantzidis.2011}
{Kazantzidis} S.,  {{\L}okas} E.~L.,  {Callegari} S.,  {Mayer} L.,
  {Moustakas} L.~A.,  2011, \mn@doi [\apj] {10.1088/0004-637X/726/2/98}, \href
  {https://ui.adsabs.harvard.edu/abs/2011ApJ...726...98K} {726, 98}

\bibitem[\protect\citeauthoryear{{Kennedy}, {Frenk}, {Cole}  \&
  {Benson}}{{Kennedy} et~al.}{2014}]{Kennedy.2014}
{Kennedy} R.,  {Frenk} C.,  {Cole} S.,   {Benson} A.,  2014, \mn@doi [\mnras]
  {10.1093/mnras/stu719}, \href
  {https://ui.adsabs.harvard.edu/abs/2014MNRAS.442.2487K} {442, 2487}

\bibitem[\protect\citeauthoryear{{Koppelman}, {Helmi}, {Massari},
  {Price-Whelan}  \& {Starkenburg}}{{Koppelman} et~al.}{2019}]{Koppelman.2019}
{Koppelman} H.~H.,  {Helmi} A.,  {Massari} D.,  {Price-Whelan} A.~M.,
  {Starkenburg} T.~K.,  2019, \mn@doi [\aap] {10.1051/0004-6361/201936738},
  \href {https://ui.adsabs.harvard.edu/abs/2019A&A...631L...9K} {631, L9}

\bibitem[\protect\citeauthoryear{{Kummer}, {Kahlhoefer}  \&
  {Schmidt-Hoberg}}{{Kummer} et~al.}{2018}]{Kummer.2018}
{Kummer} J.,  {Kahlhoefer} F.,   {Schmidt-Hoberg} K.,  2018, \mn@doi [\mnras]
  {10.1093/mnras/stx2715}, \href
  {https://ui.adsabs.harvard.edu/abs/2018MNRAS.474..388K} {474, 388}

\bibitem[\protect\citeauthoryear{{Lovell}, {Frenk}, {Eke}, {Jenkins}, {Gao}  \&
  {Theuns}}{{Lovell} et~al.}{2014}]{Lovell.2014}
{Lovell} M.~R.,  {Frenk} C.~S.,  {Eke} V.~R.,  {Jenkins} A.,  {Gao} L.,
  {Theuns} T.,  2014, \mn@doi [\mnras] {10.1093/mnras/stt2431}, \href
  {https://ui.adsabs.harvard.edu/abs/2014MNRAS.439..300L} {439, 300}

\bibitem[\protect\citeauthoryear{Ludlow et~al.,}{Ludlow
  et~al.}{2017}]{Ludlow.2017}
Ludlow A.~D.,  et~al., 2017, \mn@doi [Phys. Rev. Lett.]
  {10.1103/PhysRevLett.118.161103}, 118, 161103

\bibitem[\protect\citeauthoryear{{Majewski} et~al.,}{{Majewski}
  et~al.}{2017}]{Majewski.2017}
{Majewski} S.~R.,  et~al., 2017, \mn@doi [\aj] {10.3847/1538-3881/aa784d},
  \href {https://ui.adsabs.harvard.edu/abs/2017AJ....154...94M} {154, 94}

\bibitem[\protect\citeauthoryear{{Merritt}, {van Dokkum}, {Abraham}  \&
  {Zhang}}{{Merritt} et~al.}{2016}]{Merritt.2016}
{Merritt} A.,  {van Dokkum} P.,  {Abraham} R.,   {Zhang} J.,  2016, \mn@doi
  [\apj] {10.3847/0004-637X/830/2/62}, \href
  {https://ui.adsabs.harvard.edu/abs/2016ApJ...830...62M} {830, 62}

\bibitem[\protect\citeauthoryear{{Monachesi} et~al.,}{{Monachesi}
  et~al.}{2019}]{Monachesi.2019}
{Monachesi} A.,  et~al., 2019, \mn@doi [\mnras] {10.1093/mnras/stz538}, \href
  {https://ui.adsabs.harvard.edu/abs/2019MNRAS.485.2589M} {485, 2589}

\bibitem[\protect\citeauthoryear{{Nadler} et~al.,}{{Nadler}
  et~al.}{2021}]{Nadler.2021b}
{Nadler} E.~O.,  et~al., 2021, \mn@doi [\prl] {10.1103/PhysRevLett.126.091101},
  \href {https://ui.adsabs.harvard.edu/abs/2021PhRvL.126i1101N} {126, 091101}

\bibitem[\protect\citeauthoryear{{Naidu}, {Conroy}, {Bonaca}, {Johnson},
  {Ting}, {Caldwell}, {Zaritsky}  \& {Cargile}}{{Naidu}
  et~al.}{2020}]{Naidu.2020}
{Naidu} R.~P.,  {Conroy} C.,  {Bonaca} A.,  {Johnson} B.~D.,  {Ting} Y.-S.,
  {Caldwell} N.,  {Zaritsky} D.,   {Cargile} P.~A.,  2020, \mn@doi [\apj]
  {10.3847/1538-4357/abaef4}, \href
  {https://ui.adsabs.harvard.edu/abs/2020ApJ...901...48N} {901, 48}

\bibitem[\protect\citeauthoryear{{Navarro}, {Eke}  \& {Frenk}}{{Navarro}
  et~al.}{1996}]{Navarro.1996a}
{Navarro} J.~F.,  {Eke} V.~R.,   {Frenk} C.~S.,  1996, \mn@doi [\mnras]
  {10.1093/mnras/283.3.L72}, \href
  {https://ui.adsabs.harvard.edu/abs/1996MNRAS.283L..72N} {283, L72}

\bibitem[\protect\citeauthoryear{{Navarro}, {Frenk}  \& {White}}{{Navarro}
  et~al.}{1997}]{Navarro.1997}
{Navarro} J.~F.,  {Frenk} C.~S.,   {White} S. D.~M.,  1997, \mn@doi [\apj]
  {10.1086/304888}, \href
  {https://ui.adsabs.harvard.edu/abs/1997ApJ...490..493N} {490, 493}

\bibitem[\protect\citeauthoryear{{Newton} et~al.,}{{Newton}
  et~al.}{2021}]{Newton.2021}
{Newton} O.,  et~al., 2021, \mn@doi [\jcap] {10.1088/1475-7516/2021/08/062},
  \href {https://ui.adsabs.harvard.edu/abs/2021JCAP...08..062N} {2021, 062}

\bibitem[\protect\citeauthoryear{{Pe{\~n}arrubia}, {Benson}, {Walker},
  {Gilmore}, {McConnachie}  \& {Mayer}}{{Pe{\~n}arrubia}
  et~al.}{2010}]{Penarubia.2010}
{Pe{\~n}arrubia} J.,  {Benson} A.~J.,  {Walker} M.~G.,  {Gilmore} G.,
  {McConnachie} A.~W.,   {Mayer} L.,  2010, \mn@doi [\mnras]
  {10.1111/j.1365-2966.2010.16762.x}, \href
  {https://ui.adsabs.harvard.edu/abs/2010MNRAS.406.1290P} {406, 1290}

\bibitem[\protect\citeauthoryear{{Planck Collaboration} et~al.,}{{Planck
  Collaboration} et~al.}{2014}]{Planck_Collaboration.2014}
{Planck Collaboration} et~al., 2014, \mn@doi [\aap]
  {10.1051/0004-6361/201321529}, \href
  {https://ui.adsabs.harvard.edu/abs/2014A&A...571A...1P} {571, A1}

\bibitem[\protect\citeauthoryear{{Pontzen} \& {Governato}}{{Pontzen} \&
  {Governato}}{2012}]{Pontzen.2012}
{Pontzen} A.,  {Governato} F.,  2012, \mn@doi [\mnras]
  {10.1111/j.1365-2966.2012.20571.x}, \href
  {https://ui.adsabs.harvard.edu/abs/2012MNRAS.421.3464P} {421, 3464}

\bibitem[\protect\citeauthoryear{{Power} \& {Robotham}}{{Power} \&
  {Robotham}}{2016}]{Power.2016}
{Power} C.,  {Robotham} A.~S.~G.,  2016, \mn@doi [\apj]
  {10.3847/0004-637X/825/1/31}, \href
  {https://ui.adsabs.harvard.edu/abs/2016ApJ...825...31P} {825, 31}

\bibitem[\protect\citeauthoryear{{Purcell}, {Bullock}  \&
  {Kazantzidis}}{{Purcell} et~al.}{2010}]{Purcell.2010}
{Purcell} C.~W.,  {Bullock} J.~S.,   {Kazantzidis} S.,  2010, \mn@doi [\mnras]
  {10.1111/j.1365-2966.2010.16429.x}, \href
  {https://ui.adsabs.harvard.edu/abs/2010MNRAS.404.1711P} {404, 1711}

\bibitem[\protect\citeauthoryear{{Radburn-Smith} et~al.,}{{Radburn-Smith}
  et~al.}{2011}]{Radburn-Smith.2011}
{Radburn-Smith} D.~J.,  et~al., 2011, \mn@doi [\apjs]
  {10.1088/0067-0049/195/2/18}, \href
  {https://ui.adsabs.harvard.edu/abs/2011ApJS..195...18R} {195, 18}

\bibitem[\protect\citeauthoryear{{Read}, {Iorio}, {Agertz}  \&
  {Fraternali}}{{Read} et~al.}{2017}]{Read.2017}
{Read} J.~I.,  {Iorio} G.,  {Agertz} O.,   {Fraternali} F.,  2017, \mn@doi
  [\mnras] {10.1093/mnras/stx147}, \href
  {https://ui.adsabs.harvard.edu/abs/2017MNRAS.467.2019R} {467, 2019}

\bibitem[\protect\citeauthoryear{{Read}, {Walker}  \& {Steger}}{{Read}
  et~al.}{2018}]{Read.2018}
{Read} J.~I.,  {Walker} M.~G.,   {Steger} P.,  2018, \mn@doi [\mnras]
  {10.1093/mnras/sty2286}, \href
  {https://ui.adsabs.harvard.edu/abs/2018MNRAS.481..860R} {481, 860}

\bibitem[\protect\citeauthoryear{{Robertson}, {Massey}  \& {Eke}}{{Robertson}
  et~al.}{2017}]{Robertson.2017}
{Robertson} A.,  {Massey} R.,   {Eke} V.,  2017, \mn@doi [\mnras]
  {10.1093/mnras/stw2670}, \href
  {https://ui.adsabs.harvard.edu/abs/2017MNRAS.465..569R} {465, 569}

\bibitem[\protect\citeauthoryear{{Rodr{\'\i}guez-Torres}
  et~al.,}{{Rodr{\'\i}guez-Torres} et~al.}{2016}]{Rodriguez.2016}
{Rodr{\'\i}guez-Torres} S.~A.,  et~al., 2016, \mn@doi [\mnras]
  {10.1093/mnras/stw1014}, \href
  {https://ui.adsabs.harvard.edu/abs/2016MNRAS.460.1173R} {460, 1173}

\bibitem[\protect\citeauthoryear{{Sanderson} et~al.,}{{Sanderson}
  et~al.}{2018}]{Sanderson.2018}
{Sanderson} R.~E.,  et~al., 2018, \mn@doi [\apj] {10.3847/1538-4357/aaeb33},
  \href {https://ui.adsabs.harvard.edu/abs/2018ApJ...869...12S} {869, 12}

\bibitem[\protect\citeauthoryear{{Schaller} et~al.,}{{Schaller}
  et~al.}{2015}]{Schaller.2015}
{Schaller} M.,  et~al., 2015, \mn@doi [\mnras] {10.1093/mnras/stv1067}, \href
  {https://ui.adsabs.harvard.edu/abs/2015MNRAS.451.1247S} {451, 1247}

\bibitem[\protect\citeauthoryear{{Schaye}}{{Schaye}}{2004}]{Schaye.2004}
{Schaye} J.,  2004, \mn@doi [\apj] {10.1086/421232}, \href
  {https://ui.adsabs.harvard.edu/abs/2004ApJ...609..667S} {609, 667}

\bibitem[\protect\citeauthoryear{{Schaye} \& {Dalla Vecchia}}{{Schaye} \&
  {Dalla Vecchia}}{2008}]{Schaye.2008}
{Schaye} J.,  {Dalla Vecchia} C.,  2008, \mn@doi [\mnras]
  {10.1111/j.1365-2966.2007.12639.x}, \href
  {https://ui.adsabs.harvard.edu/abs/2008MNRAS.383.1210S} {383, 1210}

\bibitem[\protect\citeauthoryear{{Schaye} et~al.,}{{Schaye}
  et~al.}{2015}]{Schaye.2015}
{Schaye} J.,  et~al., 2015, \mn@doi [\mnras] {10.1093/mnras/stu2058}, \href
  {https://ui.adsabs.harvard.edu/abs/2015MNRAS.446..521S} {446, 521}

\bibitem[\protect\citeauthoryear{{Searle} \& {Zinn}}{{Searle} \&
  {Zinn}}{1978}]{Searle.1978}
{Searle} L.,  {Zinn} R.,  1978, \mn@doi [\apj] {10.1086/156499}, \href
  {https://ui.adsabs.harvard.edu/abs/1978ApJ...225..357S} {225, 357}

\bibitem[\protect\citeauthoryear{{Sestito} et~al.,}{{Sestito}
  et~al.}{2023}]{Sestito.2023a}
{Sestito} F.,  et~al., 2023, \mn@doi [arXiv e-prints]
  {10.48550/arXiv.2301.13214}, \href
  {https://ui.adsabs.harvard.edu/abs/2023arXiv230113214S} {p. arXiv:2301.13214}

\bibitem[\protect\citeauthoryear{Spergel \& Steinhardt}{Spergel \&
  Steinhardt}{2000}]{Spergel.2000}
Spergel D.~N.,  Steinhardt P.~J.,  2000, \mn@doi [Phys. Rev. Lett.]
  {10.1103/PhysRevLett.84.3760}, 84, 3760

\bibitem[\protect\citeauthoryear{{Springel}}{{Springel}}{2005}]{Volker.2005}
{Springel} V.,  2005, \mn@doi [\mnras] {10.1111/j.1365-2966.2005.09655.x},
  \href {https://ui.adsabs.harvard.edu/abs/2005MNRAS.364.1105S} {364, 1105}

\bibitem[\protect\citeauthoryear{{Springel}, {White}, {Tormen}  \&
  {Kauffmann}}{{Springel} et~al.}{2001}]{Springel.2001}
{Springel} V.,  {White} S. D.~M.,  {Tormen} G.,   {Kauffmann} G.,  2001,
  \mn@doi [\mnras] {10.1046/j.1365-8711.2001.04912.x}, \href
  {https://ui.adsabs.harvard.edu/abs/2001MNRAS.328..726S} {328, 726}

\bibitem[\protect\citeauthoryear{{Springel}, {Di Matteo}  \&
  {Hernquist}}{{Springel} et~al.}{2005}]{Springel.2005}
{Springel} V.,  {Di Matteo} T.,   {Hernquist} L.,  2005, \mn@doi [\mnras]
  {10.1111/j.1365-2966.2005.09238.x}, \href
  {https://ui.adsabs.harvard.edu/abs/2005MNRAS.361..776S} {361, 776}

\bibitem[\protect\citeauthoryear{{Springel}, {Frenk}  \& {White}}{{Springel}
  et~al.}{2006}]{Springel.2006}
{Springel} V.,  {Frenk} C.~S.,   {White} S. D.~M.,  2006, \mn@doi [\nat]
  {10.1038/nature04805}, \href
  {https://ui.adsabs.harvard.edu/abs/2006Natur.440.1137S} {440, 1137}

\bibitem[\protect\citeauthoryear{{Vasiliev}}{{Vasiliev}}{2019}]{Vasiliev.2019}
{Vasiliev} E.,  2019, \mn@doi [\mnras] {10.1093/mnras/sty2672}, \href
  {https://ui.adsabs.harvard.edu/abs/2019MNRAS.482.1525V} {482, 1525}

\bibitem[\protect\citeauthoryear{{Vasiliev}, {Belokurov}  \&
  {Evans}}{{Vasiliev} et~al.}{2022}]{Vasiliev.2022}
{Vasiliev} E.,  {Belokurov} V.,   {Evans} N.~W.,  2022, \mn@doi [\apj]
  {10.3847/1538-4357/ac4fbc}, \href
  {https://ui.adsabs.harvard.edu/abs/2022ApJ...926..203V} {926, 203}

\bibitem[\protect\citeauthoryear{Vogelsberger, Zavala, Cyr-Racine, Pfrommer,
  Bringmann  \& Sigurdson}{Vogelsberger et~al.}{2016}]{Vogelsberger.2015}
Vogelsberger M.,  Zavala J.,  Cyr-Racine F.-Y.,  Pfrommer C.,  Bringmann T.,
  Sigurdson K.,  2016, \mn@doi [Mon. Not. Roy. Astron. Soc.]
  {10.1093/mnras/stw1076}, 460, 1399

\bibitem[\protect\citeauthoryear{{Waller} et~al.,}{{Waller}
  et~al.}{2023}]{Waller.2023}
{Waller} F.,  et~al., 2023, \mn@doi [\mnras] {10.1093/mnras/stac3563}, \href
  {https://ui.adsabs.harvard.edu/abs/2023MNRAS.519.1349W} {519, 1349}

\bibitem[\protect\citeauthoryear{{Wang} \& {White}}{{Wang} \&
  {White}}{2007}]{Wang.2007}
{Wang} J.,  {White} S. D.~M.,  2007, \mn@doi [\mnras]
  {10.1111/j.1365-2966.2007.12053.x}, \href
  {https://ui.adsabs.harvard.edu/abs/2007MNRAS.380...93W} {380, 93}

\bibitem[\protect\citeauthoryear{{Wang}, {Bose}, {Frenk}, {Gao}, {Jenkins},
  {Springel}  \& {White}}{{Wang} et~al.}{2020}]{Wang.2020}
{Wang} J.,  {Bose} S.,  {Frenk} C.~S.,  {Gao} L.,  {Jenkins} A.,  {Springel}
  V.,   {White} S.~D.~M.,  2020, \mn@doi [\nat] {10.1038/s41586-020-2642-9},
  \href {https://ui.adsabs.harvard.edu/abs/2020Natur.585...39W} {585, 39}

\bibitem[\protect\citeauthoryear{{Wiersma}, {Schaye}  \& {Smith}}{{Wiersma}
  et~al.}{2009}]{Wiersma.2009}
{Wiersma} R. P.~C.,  {Schaye} J.,   {Smith} B.~D.,  2009, \mn@doi [\mnras]
  {10.1111/j.1365-2966.2008.14191.x}, \href
  {https://ui.adsabs.harvard.edu/abs/2009MNRAS.393...99W} {393, 99}

\makeatother
\end{thebibliography}
%%%%%%%%%%%%%%%%%%%%%%%%%%%%%%%%%%%%%%%%%%%%%%%%%%

\appendix

\section{Peak stellar mass functions}\label{stellar_mass_function_appendix}

\begin{figure}
    \centering
    \includegraphics{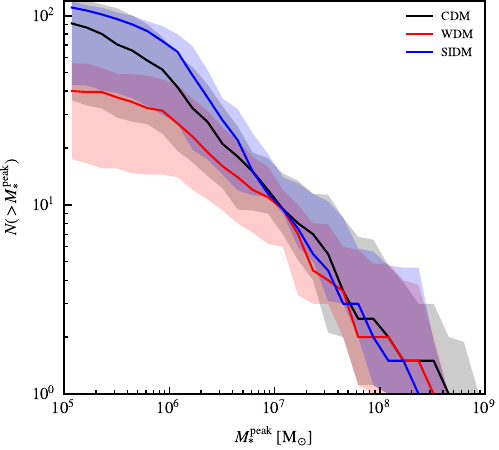}
    \caption{Similar to Fig~\ref{vpeak_distribution_orphans}, but showing instead the median peak stellar mass function of the progenitors of the stellar haloes studied in this work.}
    \label{stellar_mass_function}
\end{figure}

We compare the stellar masses of the main progenitors of the stellar haloes across models in Fig.~\ref{stellar_mass_function}, where the median peak stellar mass function is shown alongside the $1\sigma$ scatter. The peak stellar mass, $M^{\mathrm{peak}}_{*}$, is defined as the maximum bound stellar mass that the galaxy attained before being disrupted.

\section{Radial velocity vs radial distance of accreted stars}

\begin{figure}
    \centering
    \includegraphics{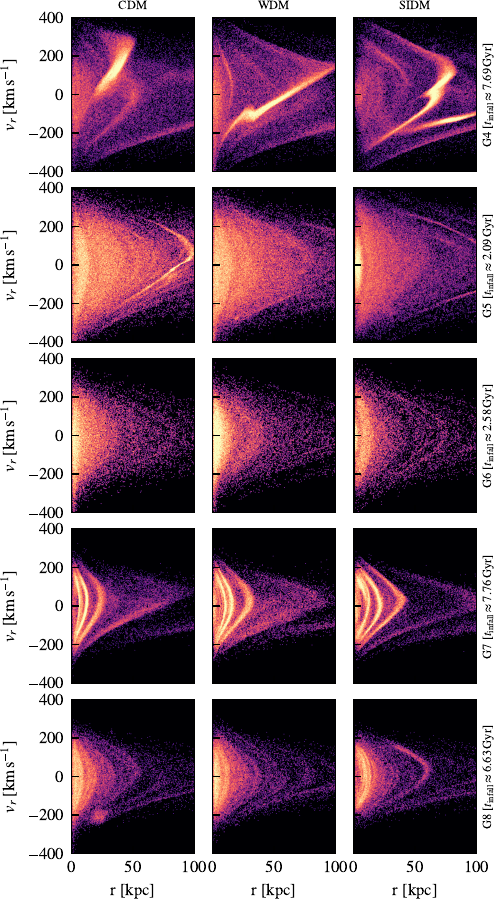}
    \caption{Radial velocity of accreted stellar particles, as a function of their radial distance to the centre of their host galaxies. The colour of each particle encodes the local ex-situ stellar density. This is shown for the remainder of the haloes not shown in Fig.~\ref{vr_vs_r_diagram_0}, with their their matched CDM, WDM and SIDM versions placed from left to right. The vertical ordering reflects the $M_{200}(z=0)$ ranking, with decreasing mass towards the lower panels. The infall time of the satellite galaxy that was involved in the last major merger for each of the galaxies in this sample is indicated on the right-hand side of the corresponding row.}
    \label{vr_vs_r_rest_of_sample}
\end{figure}

We showed in Fig.~\ref{vr_vs_r_diagram_0} how the radial velocities of accreted stars in three example stellar haloes vary with distance from their centre. This subset of the eight haloes we have in our sample was chosen for illustrative purposes in \S\ref{kinematics_stellar_halo}. For completeness, we also show the distributions of the haloes not discussed in the main text in Fig.~\ref{vr_vs_r_rest_of_sample}. 

There is a clear example of a stellar halo, that of galaxy G4, which is noticeably asymmetrical about the $v_{r} = 0$ line. The asymmetrical features are caused by the ongoing disruption of a massive satellite that was accreted at $t_{\mathrm{infall}} \approx7.69~\mathrm{Gyr}$. The merger is deemed as complete for the SIDM simulation, as the galaxy is no longer found in the galaxy catalogues. For CDM and WDM, the companion is still found as a self-bound satellite galaxy, although only a small amount of central mass remains self-bound.

Omitting the example of G4 due to the ongoing disruption of a massive satellite, we observe the same trends as those discussed in the main text. The chevrons appear in greater numbers in SIDM, and are located further out than those in CDM and WDM. However, the extend of the differences between dark matter models is highly dependent on how long ago the last major merger happened. Galaxies which had more recent major mergers have chevrons that enhance the differences found across dark matter models. The presence of chevrons is simply due to the fact that not enough time has passed for them to phase-mix into the background. In contrast, the galaxies that experienced their last major merger earlier on have largely featureless distributions of $v_{r}$ vs $r$, blurring the differences in initial chevron populations due to different dark matter models. The trend we observe between the ease of identifying differences across dark matter models based on chevrons with how long ago the last major merger happened highlights that the most constraining galaxies are likely to be those that form late.

% Don't change these lines
\bsp	% typesetting comment
\label{lastpage}
\end{document}